\documentclass[prl,reprint,showkeys,superscriptaddress]{revtex4-1}
\usepackage{graphicx}%

\begin{document}

\title{Stability of metallic edges and Fermi-level pinning in transition-metal dichalcogenide nanoribbons}
\author{Daphne Davelou}
\affiliation{Department of Materials Science and Technology, University of Crete, 71003, Heraklion, Greece.}
\affiliation{Institute of Electronic Structure and Laser, Foundation for	Research \& Technology-Hellas, N. Plastira 100,
	70013 Heraklion, Crete, Greece}
\email{d.davelou@materials.uoc.gr}

\author{Georgios Kopidakis}
\affiliation{Department of Materials Science and Technology, University of Crete, 71003, Heraklion, Greece.}
\affiliation{Institute of Electronic Structure and Laser, Foundation for	Research \& Technology-Hellas, N. Plastira 100,
	70013 Heraklion, Crete, Greece}

\author{Efthimios Kaxiras}
\affiliation{Department of Physics and Harvard's Paulson School of Engineering and Applied Sciences, Harvard University, Cambridge MA 02138, USA.}

\author{Ioannis N. Remediakis}
\affiliation{Department of Materials Science and Technology, University of Crete, 71003, Heraklion, Greece.}
\affiliation{Institute of Electronic Structure and Laser, Foundation for	Research \& Technology-Hellas, N. Plastira 100,
	70013 Heraklion, Crete, Greece}


\begin{abstract}
Nanoribbons of MoS$_2$ present a unique electronic structure that consists of a semiconducting bulk bounded by metallic edges; same holds for other Transition-Metal Dichalcogenides (TMDs)   (Mo-,W-,S$_{2}$,Se$_{2}$).
 We perform first-principles calculations for TMD nanoribbons with reconstructed zig-zag metal terminated edges that contain chalcogen adatoms. 
All nanorobbons have possitive edge energies when the chemical potential of chalcogens is close to the energy of solids, and negative edge energies for high chemical potential. 
The reconstruction with two chalcogen adatoms is expected to be the most stable one. In all nanoribbons, a metallic phase is found near their edges, with the Fermi level of this metallic phase being lower than the Fermi level of the 2D material. 
\end{abstract}

\keywords{Transition metal dichalcogenides; TMD nanoribbons; DFT- first principal calculations; Edge states}

\maketitle

\section{Introduction}
Graphene-like layered materials are extensively studied for their unique mechanical, electronic and chemical properties. Popular among them are the transition metal dichalcogenides (TMDs) of the type MX2, where M is a metal and X is chalcogen \cite{wang12}. Such materials include MoS$_2$, MoSe$_2$, WS$_2$ and WSe$_2$ which are semiconductors in both their 3D and 2D structure \cite{mak10,splendiani10,kuc11,ding11,liu11,kadantsev12,tritsaris13,santos13,maniadaki16a}, while new members are constantly added to this family \cite{lebegue2013}. Although semiconducting in 2D, there is strong evidence that MoS$_2$ is metallic in its quasi 1D structures \cite{helveg00,bollinger01,bollinger03}. 

 Apart from using them as a model system for the structure of edges, MoS$_2$ nanoribbons have been studied extensively as they offer a unique electronic structure that combines features of MoS$_2$ and graphene. Bollinger {et al.} \cite{bollinger01,bollinger03} were the first who predicted metallic edge states on MoS$_2$ nanoribbons using DFT simulations. Later, such nanoribbons were synthesized by  Camacho-Bragado {em et al.} attached to MoO$_3$ clusters \cite{CamachoBragado05} and by Wang {\em et al.} inside carbon nanotubes \cite{wang10}. Several theoreticians have simulated such nanoribbons: Li {\em et al.} \cite{li08} predicted that zig-zag MoS$_2$ nanoribbons are magnetic and are more stable than triangular nanoclusters. Kou  {\em et al.} \cite{kou12} showed that strain and applied electric field can alter dramatically both magnetism and electronic structure. In a previous work \cite{davelou14}, we had found that the metallic edges alter the dielectric permitivity of the nanoribbons and that their spatial extent is around 5 \AA\ from the edge of the material.
Kim {\em et al.} \cite{kim15} found that armchair nanoribbons can be stabilized by H adsorption and possess a semiconducting character with strong exciton effects. Finally, Mos$_2$ nanoribbons could have a key role in the hydrogen evolution reaction \cite{hinnemann05,hinnemann05b}, and graphene support enhances these features \cite{maniadaki16b,Tsai14}. Yu  {\em et al.} \cite{yu16} verified the tunability of edge states using electric field and hydrogen absorption.

 Although the presence of metallic states in MoS$_2$ and their electronic structure is well established, two questions remain open: First, what is the stability of such edges in relevant experimental conditions. The limited number of reports for their synthesis implies that such structures might be unfavorable, while their use in catalysis confirms that, once formed, these structures are extremely stable. Second, it is not clear whether the metallic edges are favorable or not for electrons of holes of the semiconducting bulk. In other words, what is the position of Fermi level of the metallic region relative to the Fermi level  of the bulk. 

 In this work, we attempt to provide answers to both these questions. We start by verifying the presence of metallic edges in MoS$_2$ and other TMDs, with different edge terminations. We calculate the stability of edges using the chemical potential of chalcogens as a free parameter. Finally, we align the band structure and DOS of the nanoribbon with that of the two-dimensional single layer and discuss band offsets and changes in the Fermi level. 

\section{Computational method}
\subsection{DFT implementation}
We perform DFT  calculations using the open-source Grid-based Projector Augmented-Wave (GPAW) package  \cite{GPAW, Enkovaara}. 
GPAW employs the PAW method \cite{Bloch} and uses real-space grids for the representation of the wave functions and the electron densities. For the calculation of the exchange-correlation functional we use the Generalized Gradient Approximation (GGA) of Perdew-Burke-Ernzerhof (PBE) \cite{Perdew}. Although PBE functional is not the preferable choice for the calculation of the band structure due to its underestimation of the energy gap \cite{qiu13}, the use of hybrid functionals and/or many-body equations to treat accurately the excitonic effects \cite{Ramasub12,bhimanapati15,qiu16} is well beyond the scope of this work  where we mostly focus on ground-state properties or trends for the position of the Fermi level.

 The Atomic Simulation Environment (ASE) open-source suite \cite{ASE} is used for the generation of structures, atomic relaxation and analysis of the results.

 In GPAW, the computational parameters that need to be taken into account are the grid spacing, $h$, the number of Monkhorst-Pack $\mathbf{k}$-points in the Brillouin zone and the thickness of the vacuum region beyond the last atoms of the system. The parameters used in our study are $h\approx{0.19}$\AA, $4\mathtt{x}1\mathtt{x}1$ $\mathbf{k}$-points and $L_V=12.0$\AA. For each $\mathbf{k}$-point we use 344 eigenstates. This number of bands corresponds to $N/2+200$ states for $n_c=6$, where $N$ is the number of valence electrons.
 
\subsection{Model structures}
We create TMD nanoribbons of the four most common two-dimensional TMDs: MoS$_2$, MoSe$_2$, WS$_2$ and WSe$_2$.  For each material, we start from a rectangular unit cell with sides $a$ along $x$ and $a\sqrt{3}$ along $y$, where $a$ is the theoretical lattice constant of the 2D material. We use the theoretical lattice parameters for the TMDs as found in Ref. \cite{maniadaki16a}. The unit cell contains 2 metal and 4 chalcogen atoms. We construct a supercell by repeating this unit cell $n_c$ times along $y$ and then a vacuum region is added along $y$ and $z$. Periodic boundary conditions are imposed along all axes. The structure formed is a nanoribbon of infinite length along $x$ and has a width of $n_c$ cells along $y$, with ($1 \le n_c \le 7$). This corresponds to nanoribbon widths in  the range of 3.5\AA\  to 42.4\AA. 
 
\begin{figure}
	\includegraphics[width=0.175\linewidth]{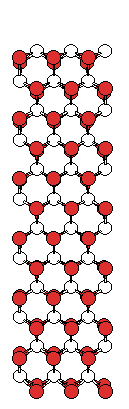}
	\includegraphics[width=0.18\linewidth]{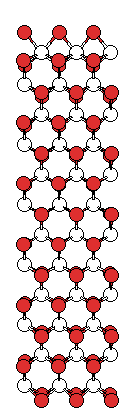}
	\includegraphics[width=0.20\linewidth]{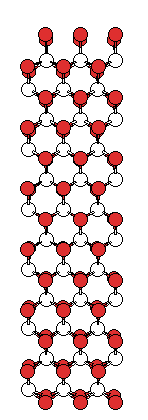}
	\includegraphics[width=0.19\linewidth]{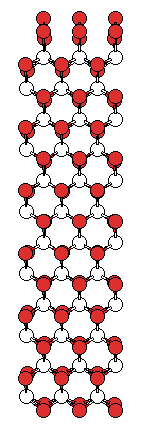}
	\includegraphics[width=0.18\linewidth]{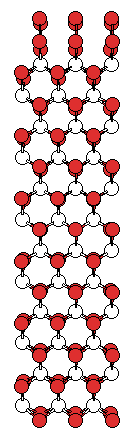}
	\caption{\label{1}Ball-and stick model of relaxed structures of typical nanoribbons (MoSe$_2$ with $n_c=6$) with increasing number of adatoms , from zero (left) to four (right), at the metal-terminated zig zag edge. Three unit cells of the periodic structure are shown. White and red spheres represent the metal and chalcogen atoms, respectively.}
\end{figure}

 For every width, $n_c$, we consider five different terminations of the M-edge. These terminations correspond to $N_X$  ($0 \le N_X \le 4$) X adatoms decorating the M-edge (see Fig. \ref{1}). In total, we consider 140 nanoribbons (four materials times seven widths times five terminations).  For each nanoribbon, we allow full relaxation of atom positions in the first two layers from both edges, as well as all adatom positions.
\subsection{Edge energies}
\paragraph{}
The edge energy, $\gamma$, is defined in a similar way as the surface energy of three-dimensional materials: A nanoribbon made of MX$_2$ has total energy 

\begin{equation}
E_{tot} = 2n_cE_{MX2}+N_X\mu_X + 2a\gamma
\end{equation}
where  $E_{MX2}$ is the energy of the monolayer per MX$_2$ unit, $2n(c)$ is the number of MX$_2$ units along the $y$ axis, $N_{X}$ is the number of adatoms at the reconstructed edge and $\mu_{X}$ is the chemical potential of the chalcogen X. The last term is the energy cost associated with the formation of edges, which is proportional to the length of the unit cell, along $x$, $a$, and the  edge energy, $\gamma$. The edge energy is calculated by plotting $E_{tot}$ as a function of $n_c$ and then using least-square method to fit a straight line to the calculations. More details can be found in Ref. \cite{davelou14}.
 
 The chemical potential of chalcogen X (S or Se) is an important parameter that links the simulation to the experimental conditions. In principle, the nanoribbons are at equilibrium to a reservoir of X atoms, and $\mu_X$ is the energy required to take one X atom from this reservoir. No matter what the X-containing compound is, the energy of X cannot be greater than the energy of an X atom (where X has no chemical bonds), and it cannot be lower than the energy of solid X, which is the preferred state of X at standard conditions.
 
 In this study, we performed  DFT calculations at the same level of accuracy for TMD nanoribbons, chalcogen atoms and solid chalcogens, in order to obtain the total energy per atom in these structures.

\begin{figure*}
	\begin{tabular}{cc}
		\includegraphics[width=0.5\linewidth]{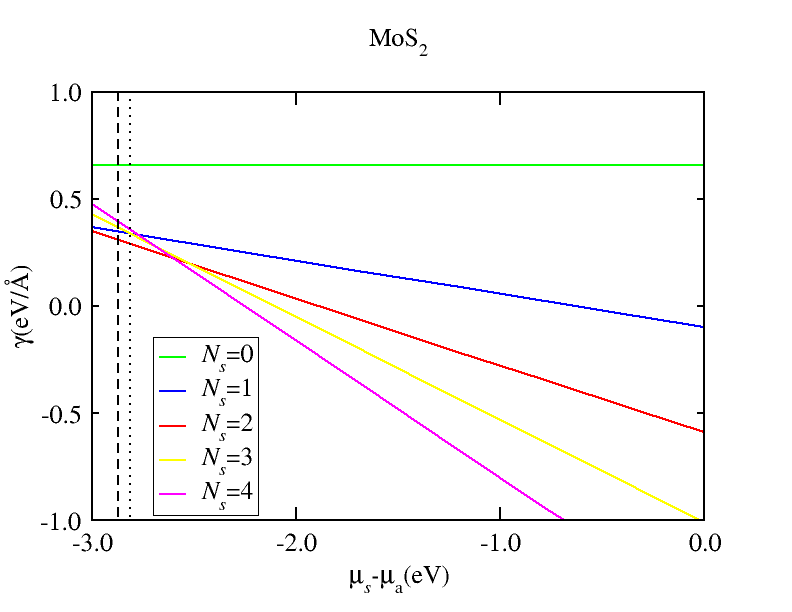} &
		\includegraphics[width=0.5\linewidth]{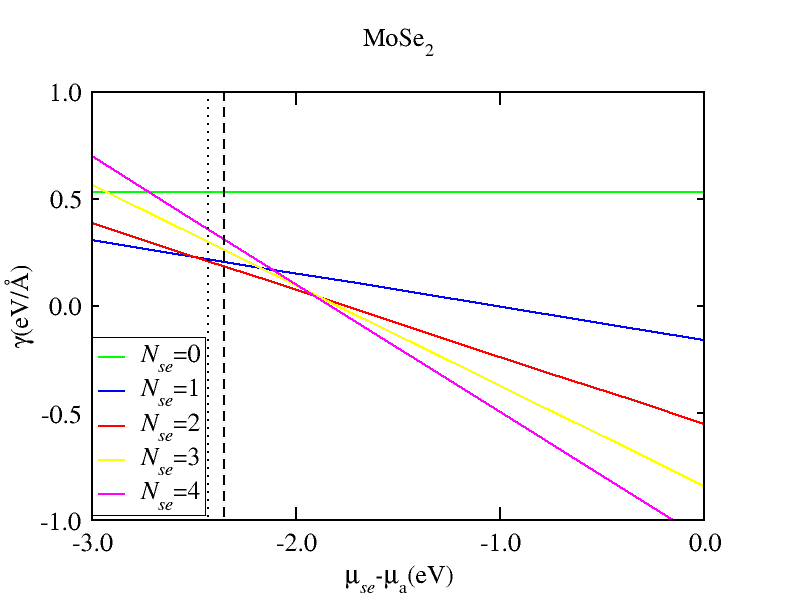} \\
		\includegraphics[width=0.5\linewidth]{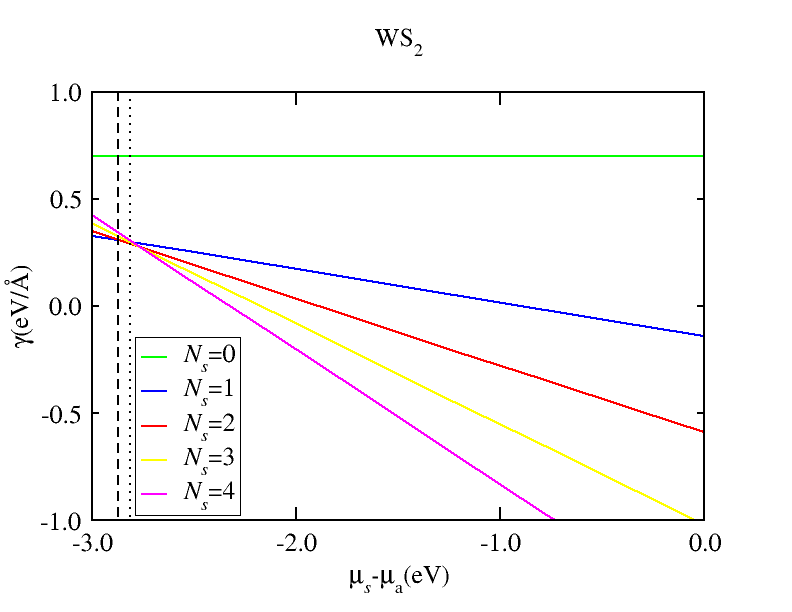} &
		\includegraphics[width=0.5\linewidth]{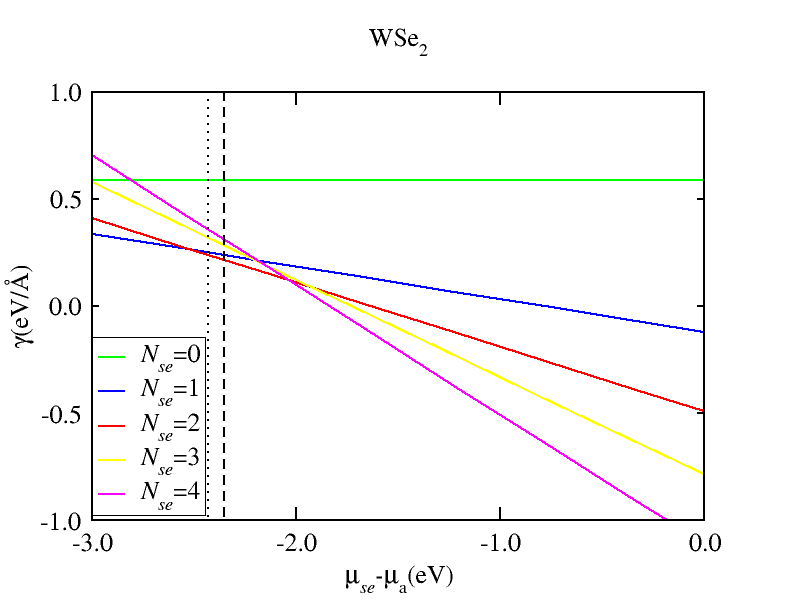}
	\end{tabular}
	\caption{\label{2}Edge energies, $\gamma$, of MX$_2$ nanoribbons with different number of adatoms, as a function of the chalcogen chemical potential, $\mu_X$ minus   is the chemical potential of atomic X, $\mu_a$. Different colours correspond to different number of adatoms. Dotted (dashed) lines are drawn at the theoretical (experimental) cohesive energy of solid X.}
\end{figure*}

\section{Results and Discussion}
\subsection{Structural Properties}
\subsubsection{Cohesive energy of S and Se}
Chalcogens have perhaps the most complicated crystal structures \cite{wyckoff} found for elemental solids. A full DFT study of these structures is a difficult project by itself and is well beyond the scope of this work. We limited ourselves to calculations of metastable structures that are  simpler than the standard forms of the materials, while maintaining the same local environment around each chalcogen atom.  We use lattice parameters from \cite{wyckoff} and do not allow  relaxation of atomic coordinates or unit cell parameters. Both chalcogens prefer to form structures with two-fold coordination, where rings of atoms are arranged periodically in space. The rhombohedral structure of S is modelled using a hexagonal unit cell that contains three rings of six S atoms. The structure of $\alpha$-monoclinic Se has a unit cell that contains four eight-membered rings. 

 As a first validation of our calculations, we considered the cohesive energy of S and Se, which is the difference in energy between an isolated atom and an atom in the solid. We compare these values to the atomization energies of these materials \cite{webelements}. The calculations are in excellent agreement to experimental data. The cohesive energy of S is found to be 2.82 eV (experimental value is 2.87 eV) while the one of Se is found to be 2.43 eV (experimental value is 2.35 eV).

\subsubsection{Edge energy of TMD nanoribbons}
The relaxed structures of five typical MX$_2$ nanoribbons are shown in Fig. \ref{1}. In all cases, chalcogen adatoms are two-fold coordinated and have similar bond lengths as in the bulk of the material. 

 The edge energy, $\gamma$, which is the one-dimensional analogue of surface energy/surface tension, is the energy per unit length that is needed in order to cleave the material. A 2D material in vacuum is expected to have $\gamma > 0$, as it costs energy to break chemical bonds and create an edge. For a 2D material in equilibrium with an active compound, the edge energy will be different, as new bonds can be formed between edge atoms and atoms of molecules coming from the reservoir. By applying a general formula for the solid-gas interfacial tension \cite{barmparis12,barmparis15}, we can write that 

\begin{equation}
\gamma = \gamma_0 - E/L,
\label{2}
\end{equation}
where $\gamma_0$ is the edge energy in vacuum, $E$ is the average energy of any new bonds that are formed and $L$ the average distance between such new bonds. 

 From equation (\ref{2}), it is clear that $\gamma$ could be either negative or positive, depending on the strength and density of the bonds between edge atoms and atoms/molecules of the reservoir. A negative value of $\gamma$ means that the ripping of the material and formation of edges is an exothermic process, whereas ripping of the material is endothermic for $\gamma >0$. 

 The results for the dependence of edge energy on the chemical potential of chalcogens are shown in Fig. \ref{2}. For convenience, we give chemical potential with reference to the chemical potential of a chalcogen atom at zero pressure and temperature, $\mu_a$ and also we mark the chemical potential of the respective elemental solid by vertical lines. We advocate that the experimentally relevant region of chemical potentials is close to- and to the right of these lines. In chemical compounds of chalcogens (such as alkanethiols), X forms the same number of covalent bonds as in solid X, and should have similar energy (although a bit higher than in its standard form). 

 Depending on the value of the edge energy, we expect that when a single layer of a transition metal dichalcogenide is exposed to a chalcogen environment, it will either remain intact (positive $\gamma$ ) or it will spontaneously split into ribbons (negative $\gamma$ ). In all four materials,  $\gamma$ is negative for $\mu \sim \mu_a$, as X atoms can lower their energy by leaving the gas phase and adsorbing to the TMD edge. For this reason, the structure with the largest possible number of adsorbed X atoms (in our study this corresponds to $N_X=4$) is always favored for high values of $\mu$. Moreover, $\gamma$ is negative in all cases when the chemical potential is just half an eV above the energy of the solid. This means that edges will be formed spontaneously if the 2D material is at equilibrium with atomic X or even X compounds where X is weakly bound. On the other hand, $\gamma$ is positive  for chalcogen chemical potential near that of the solid S, which means that TMD edges will be quite stable with respect to most sulfides and selenides, and are vulnerable to ripping only when in contact with atomic chalcogens or radicals. 
 
 The plots of Fig. \ref{2} can also provide the favourable number of adatoms for each material, which is the number that corresponds to the line that is lowest in energy than the others. The Mo,W-terminated edge has energy that is significantly higher (at least by about 0.3 eV/{\AA}) than the energy of any other edge. 
However, metal-terminated  MoS$_2$, MoSe$_2$, WS$_2$ and WSe$_2$ will form stable 1D structures by attracting atoms at the zig-zag edge, once they are exposed to an external source of chalcogens. In the case of solid chalcogen reservoir (dashed line at low $\mu$), the preferable number of adatoms at the reconstructed edge is two, while for higher values of the chemical potential, the preferred amount of adatoms in the edge increases, as expected.In the case of atomic source, the edge energies of the nanoribbons
drops by 0.3eV/{\AA} to 0.4eV/{\AA} per extra adatom. The number of chalcogen atoms bound to the nanoribbons edge could be even higher than four. 

 In all materials, the edge with two adatoms has the lowest energy for chalcogen chemical potential close to those of  solid S and Se. For MoS$_2$ in particular, our findings are in excellent agreement to experiments that found complete absence of M edges and preference for edges terminated by S dimers , in agreement with experimental observations \cite{CamachoBragado05}.

\begin{figure*}
	\begin{tabular}{cc}
	\includegraphics[clip=true, trim=0.5cm 1cm 1cm 4cm, width=0.5\linewidth]{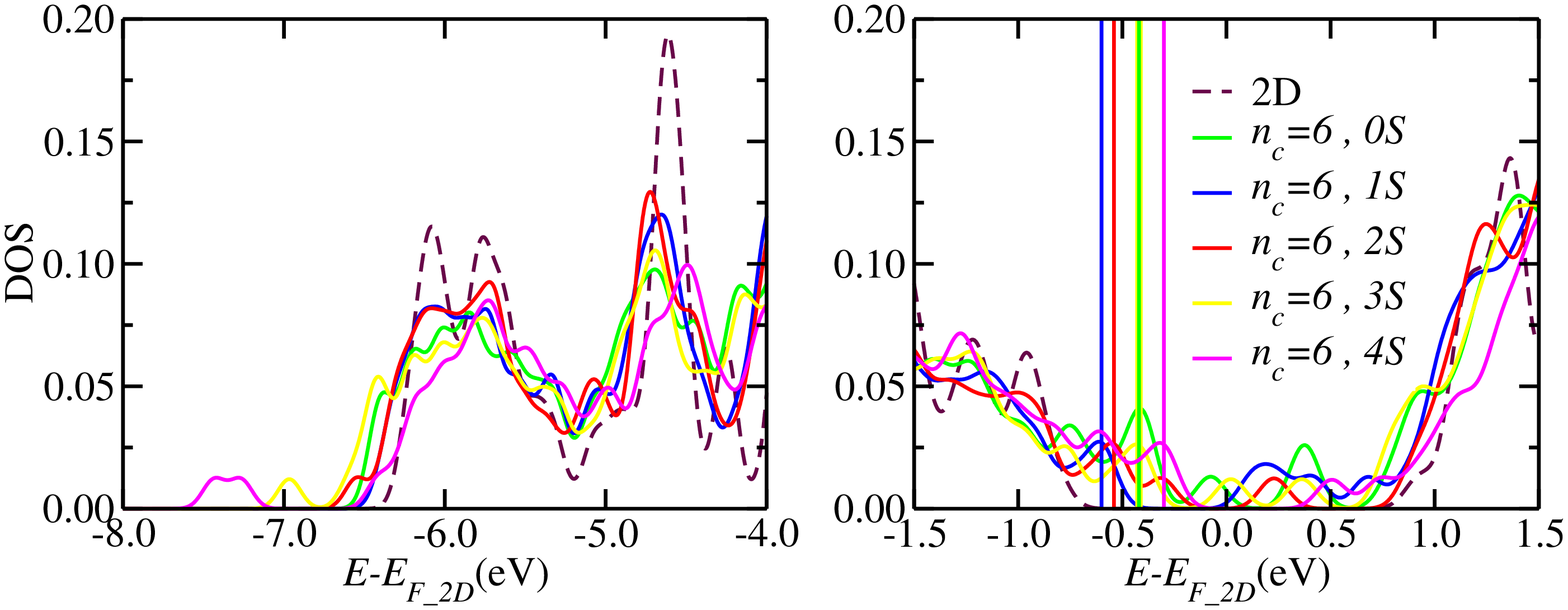} &
	\includegraphics[clip=true, trim=0.5cm 1cm 1cm 4cm, width=0.5\linewidth]{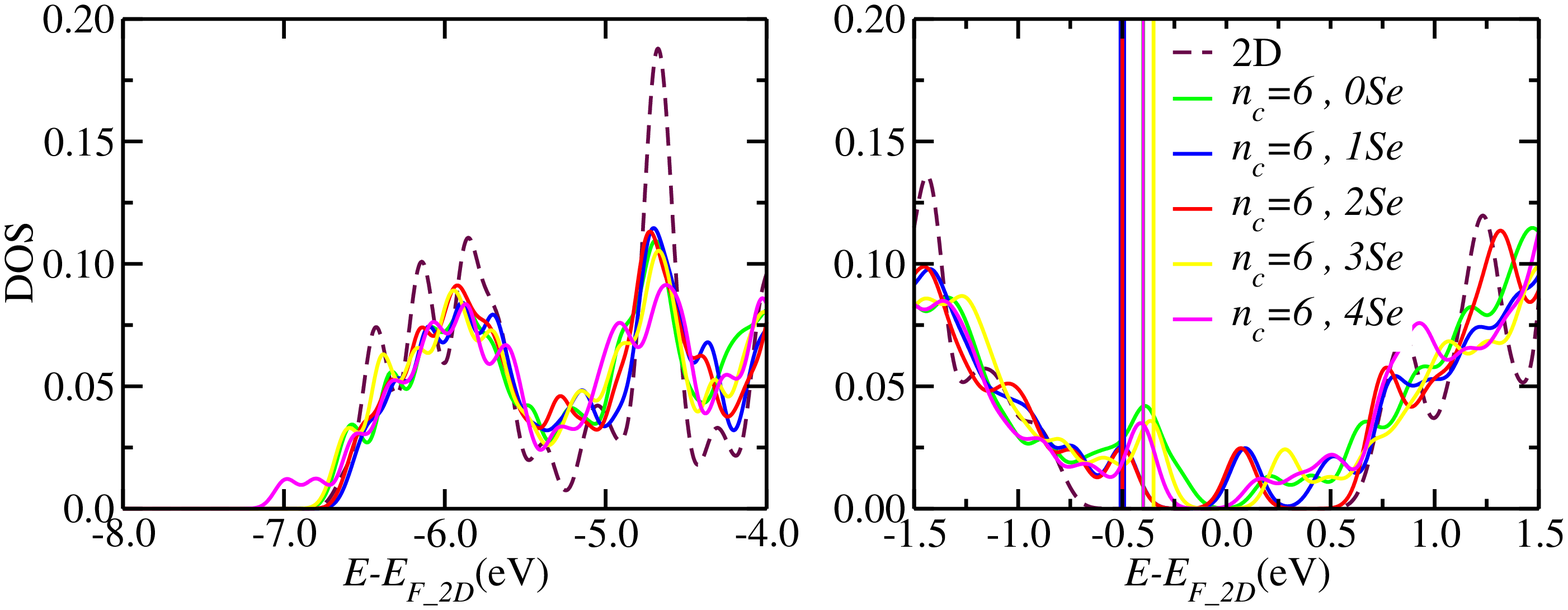} \\
	\includegraphics[clip=true, trim=0.5cm 1cm 1cm 4cm,, width=0.5\linewidth]{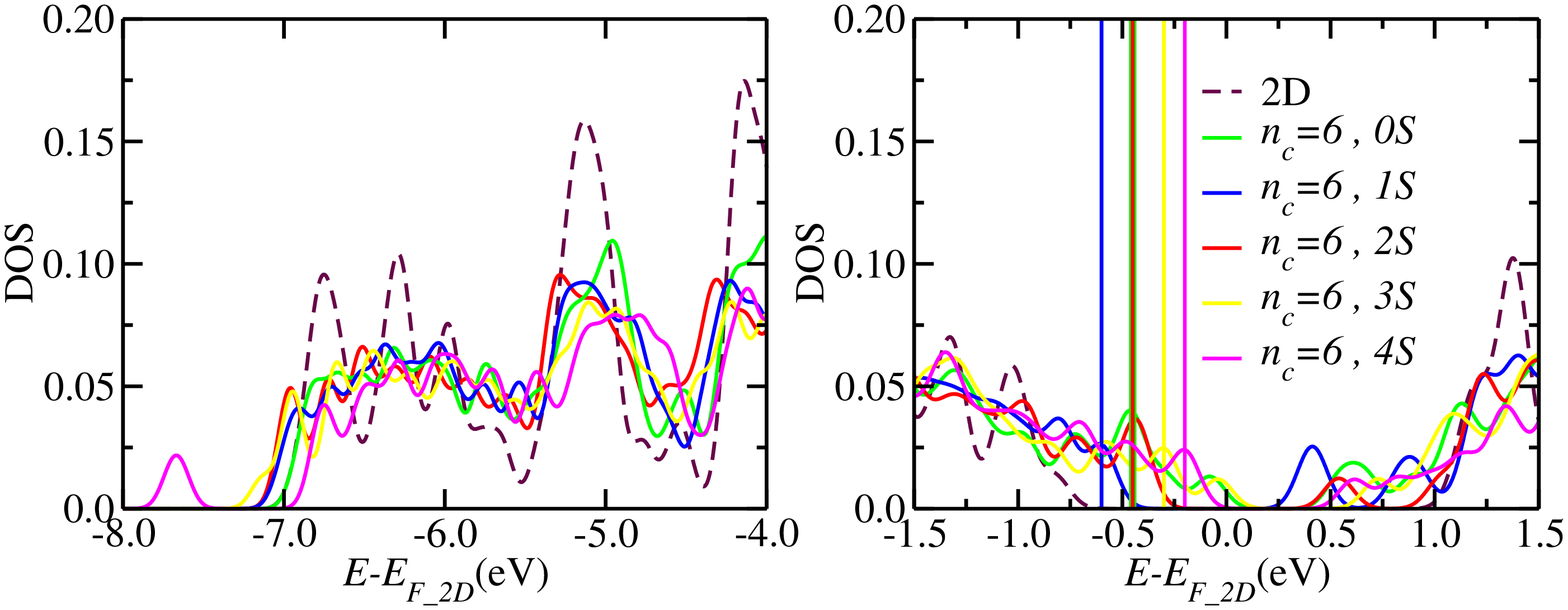} &
	\includegraphics[clip=true, trim=0.5cm 1cm 1cm 4cm,, width=0.5\linewidth]{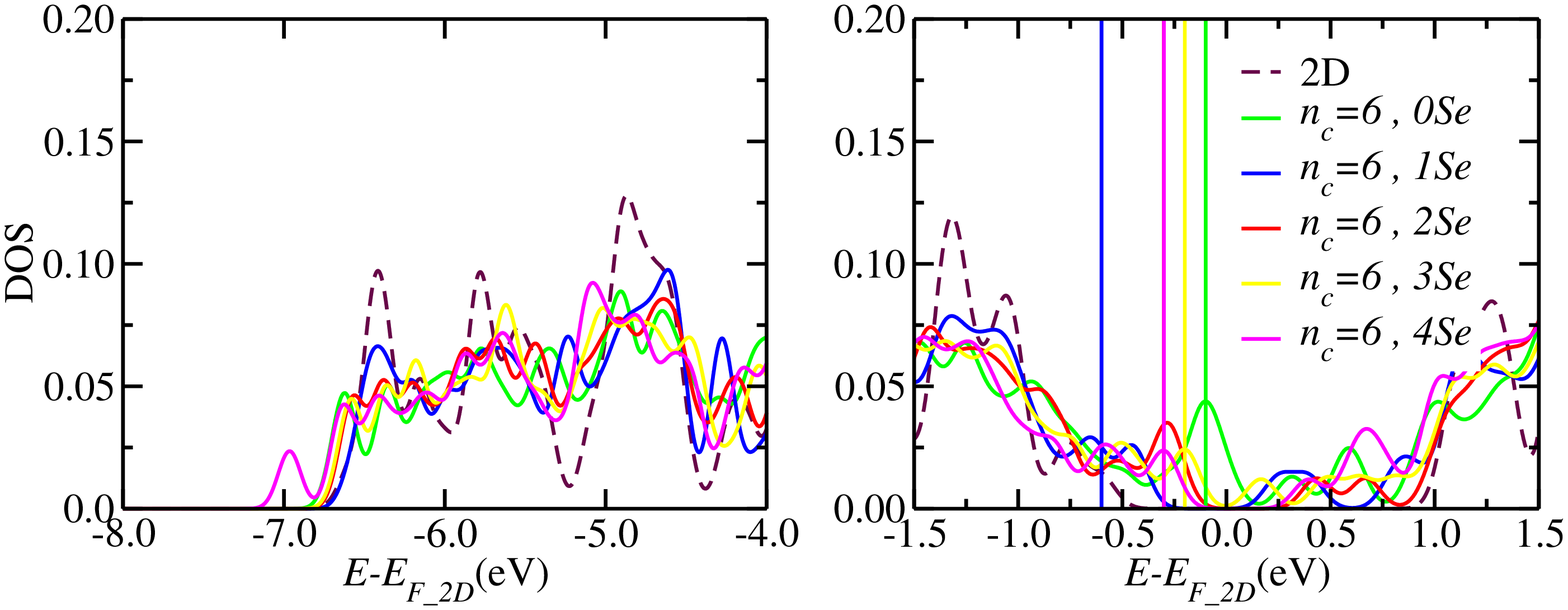}
		\end{tabular}
	\caption{\label{3}DOS for 2D MX2 (dashed lines) and the MX2 nanoribbons (coloured solid lines, see legends). For each material, we present zooms of the DOS plot at low energies (left) and close to the gap of the 2D material (right). Energies are measured with respect to the Fermi level of the 2D material, where Fermi levels $E_F$ of the nanoribbons are shown with vertical lines. From top to bottom: MoS$_2$, MoSe$_2$, WS$_2$ and WSe$_2$.}
\end{figure*}

\subsection{Electronic properties}
The density of states (DOS) of a TMD nanorribon includes a characteristic peak at the Fermi level, $E_F$, surrounded by few lower peaks that also lie inside the gap of the 2D material. These states are localized at the upper and lower edge of the nanoribbon along the $y$ axis and resemble electrons confined into one-dimensional motion. The height of the peak at $E_F$ decreases with increasing width, $n_c$, of the nanoribbon. At $n_c\to\infty$, the peak is expected to disappear. The reason for this width-dependence is the following: In the valence band of a nanoribbon with width $n_c$ and $N$ adatoms there are $18n_c+3N$ occupied electronic states  for each $\mathbf{k}$-point in the Brillouin zone, only few of which are edge states. The percentage of edge states becomes negligible for large $n_c$.  
 
 We observe exactly the same behaviour in all TMDs, all widths and all numbers of chalcogen adatoms: there is a clear peak at $E_F$, the height of which decreases for increasing nanoribbon width. In all cases, the states that contribute to this peak at the Fermi level are edge states. The DOS of all 140 nanoribbons considered in this study is presented in the Fig. \ref{6} of the Appendix.  

 In the following, we present the electronic structure of nanoribbons with width  $n_c=6$, similar to those shown in Fig. \ref{1}. At this width, there are enough bulk states to represent the DOS of the 2D material, while the edge states are prominent and have impact on the properties of the system.

 Dimensionality is a key parameter in the electronic structure of TMDs. In 3D structures, all materials considered in this study are indirect gap semiconductors \cite{kam82}. The 2D single-layer structure is a direct gap semiconductor, with strong excitonic effects \cite{kumar12}. This shift of the energy gap from $K-\Gamma$ to $K-K$ points of the Brillouin zone is responsible for the observed  photoluminescence. The energy gap of the 2D materials disappears in the 1D structures (nanorribons).

\begin{figure*}
	\includegraphics[scale=0.22]{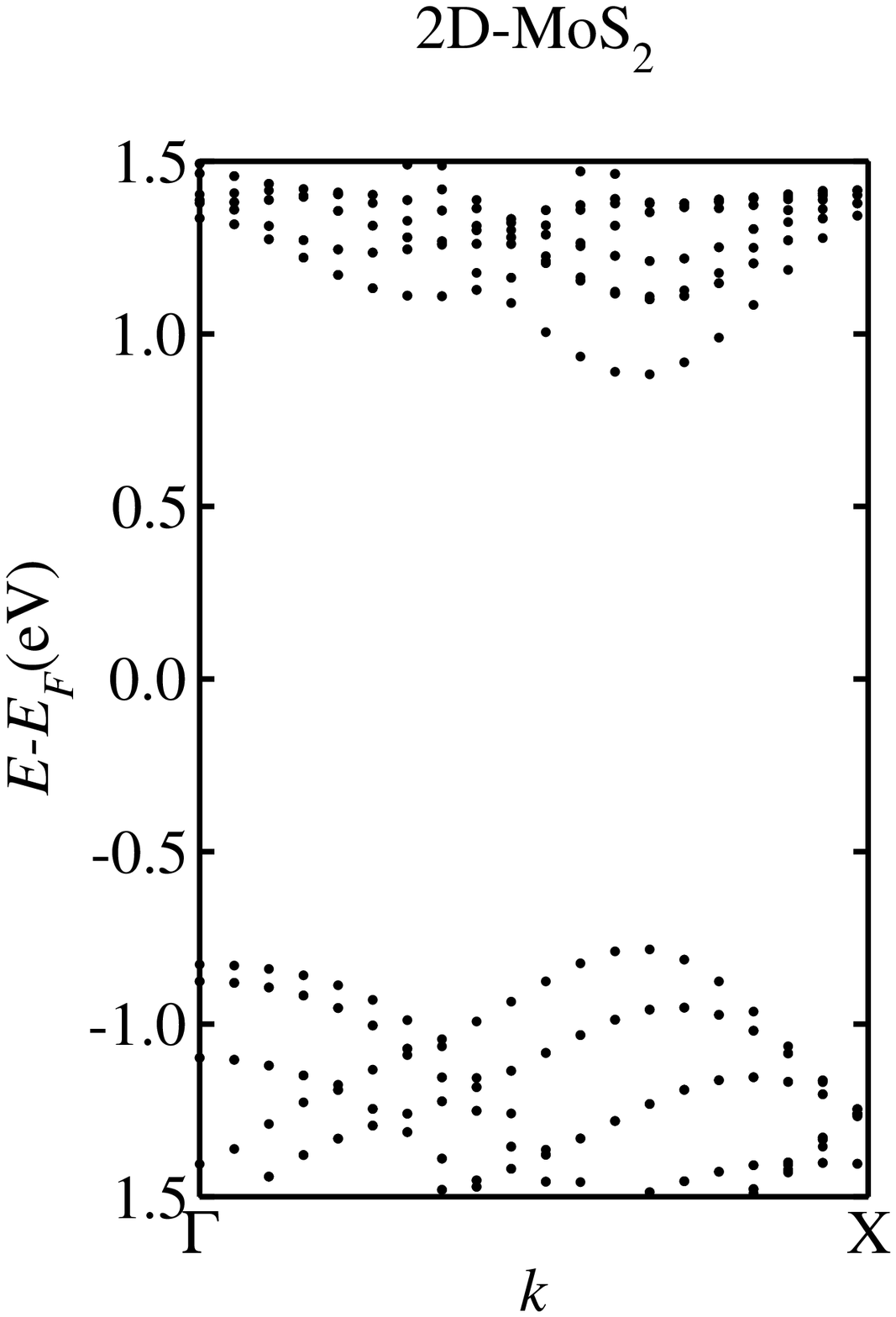}\includegraphics[scale=0.22]{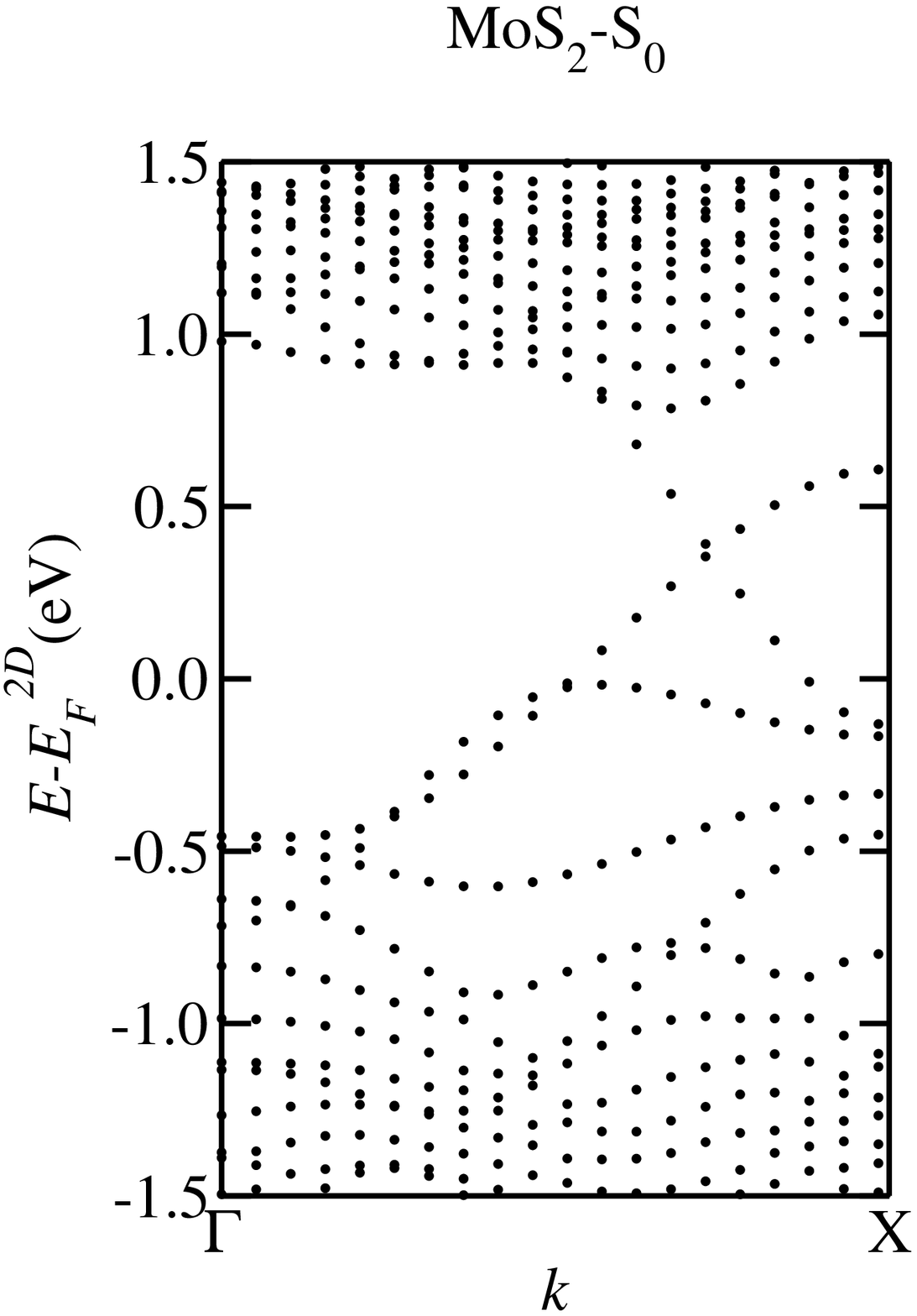}\includegraphics[scale=0.22]{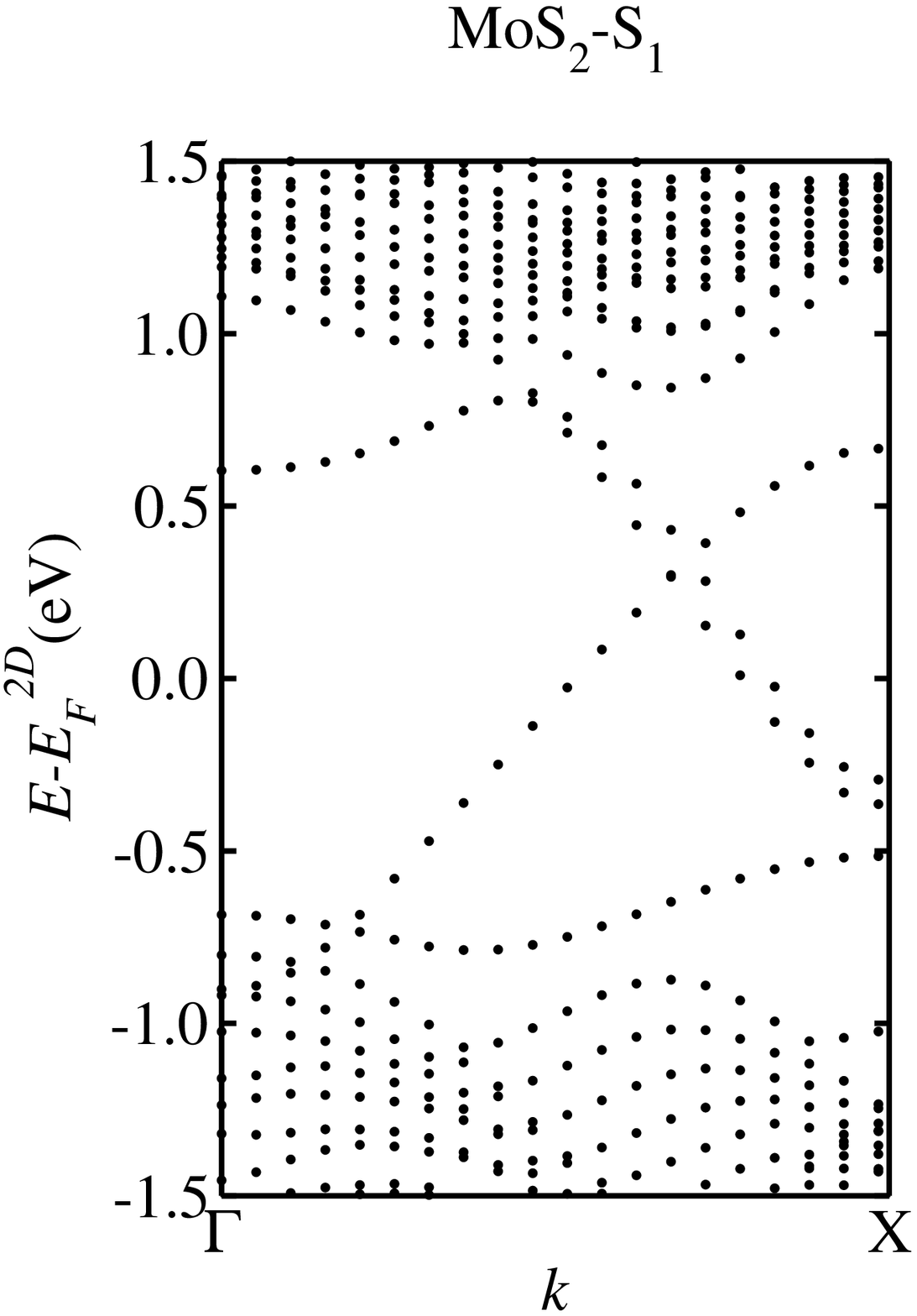}\includegraphics[scale=0.22]{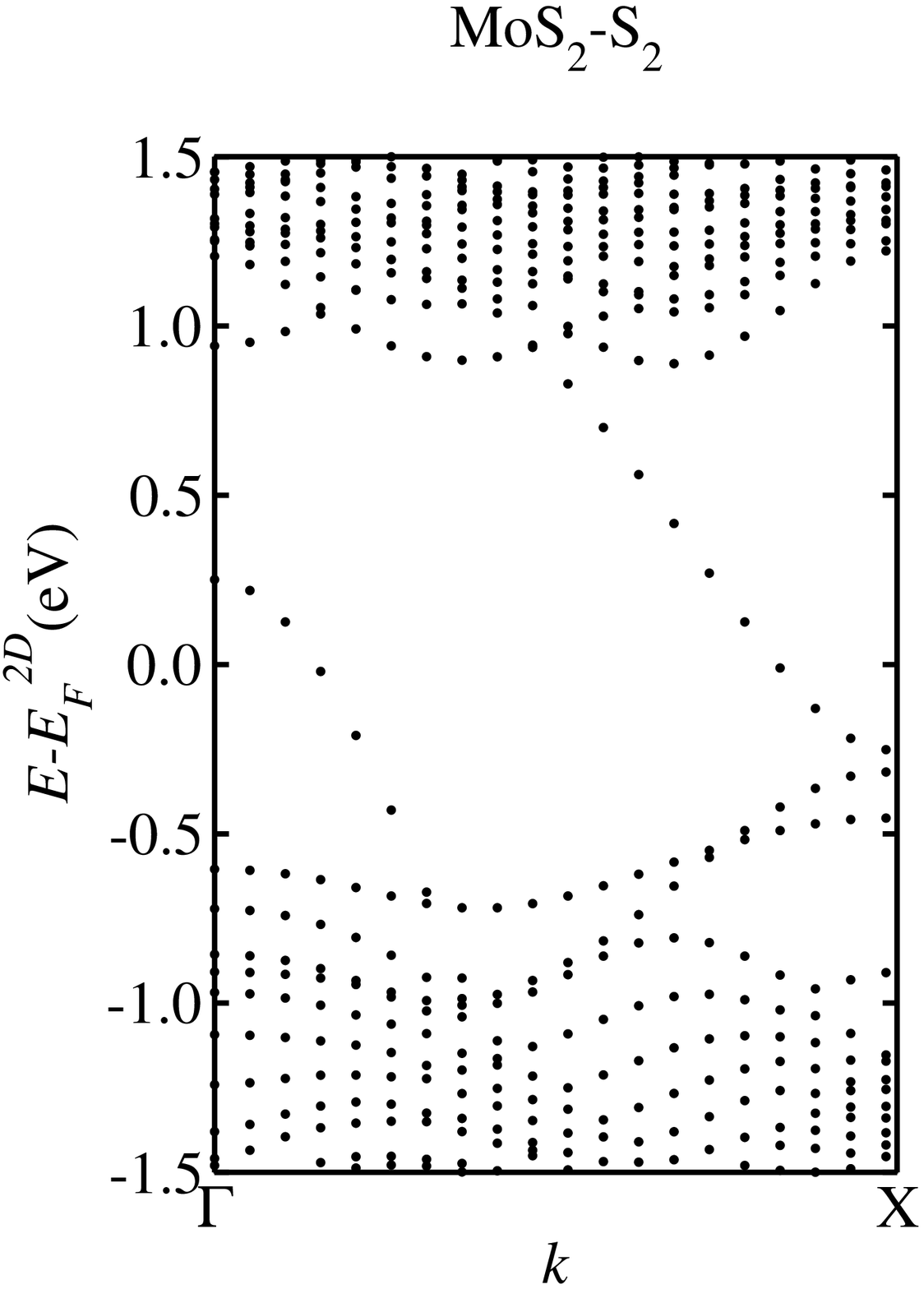}\includegraphics[scale=0.22]{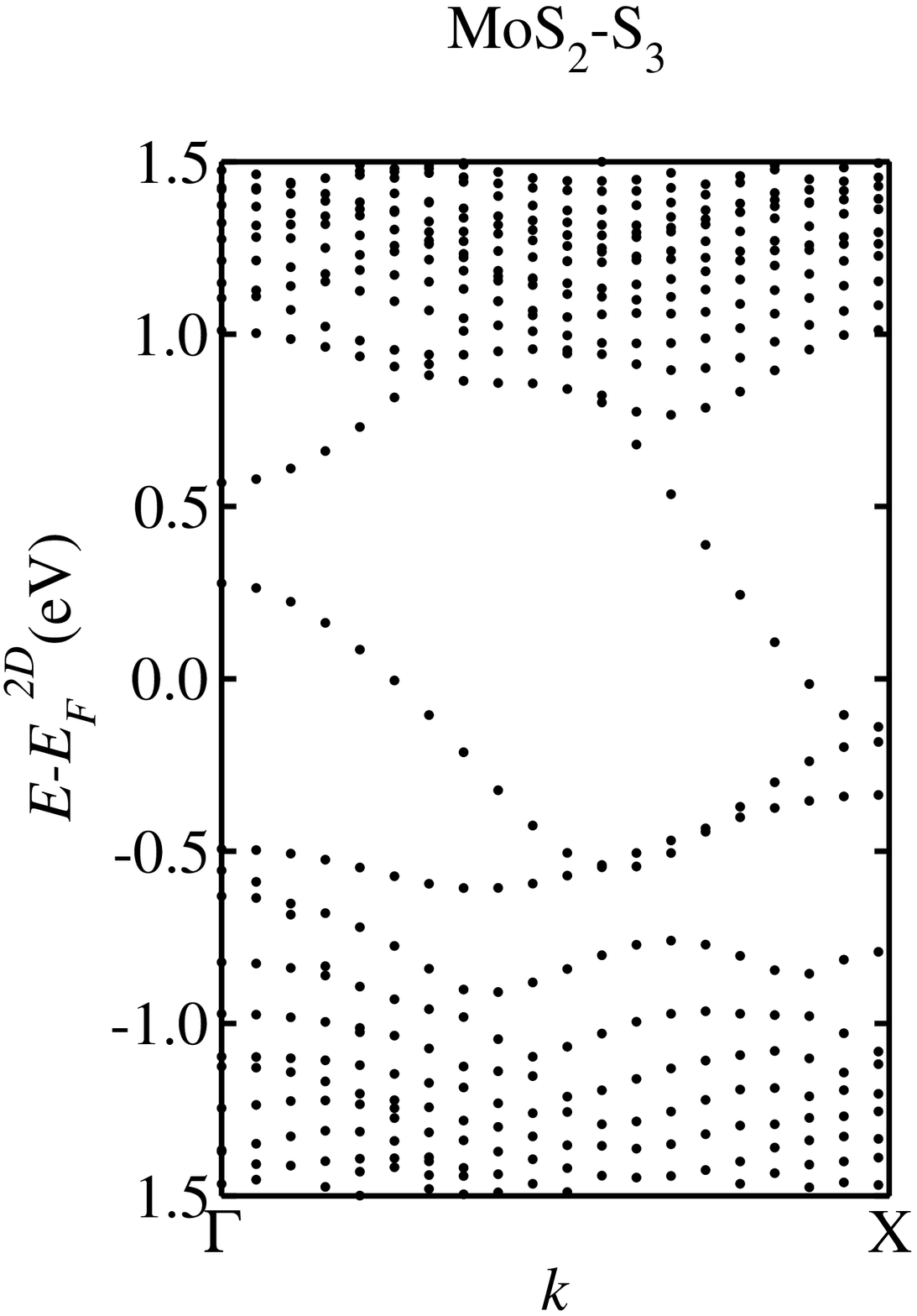}\includegraphics[scale=0.22]{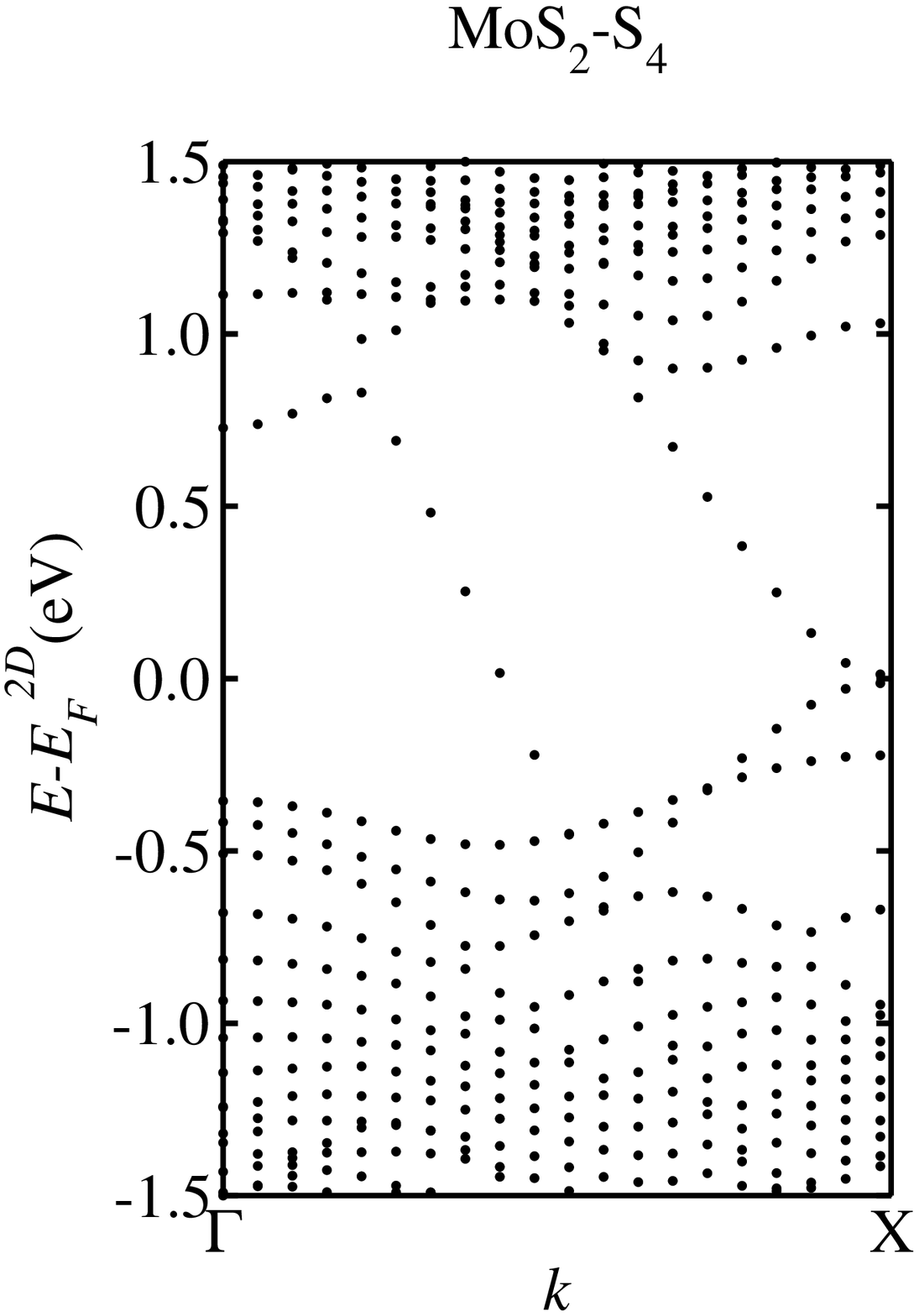}
	\caption{\label{5}Band-structure diagrams of the MoS$_2$ nanoribbon with width $n_c=6$ and different number of S adatoms along the Mo zig-zag edge. $E_F^{2D}$ is the Fermi energy of the single-layer MoS$_2$.}
\end{figure*}

\subsubsection{Density of states of TMD nanoribbons}
The DOS alone might not be enough to characterize the edge states as metallic, as it is important to verify that indeed the Fermi level of the nanoribbon lies within the gap of the 2D material. Aligning the DOS of two different systems in order to calculate the difference of their Fermi levels is a non-trivial task. Here, we adopt a simple procedure which is based on the assumption that states at the lowest edge of the valence band are not affected by the presence of edge in the material.

 Our DOS calculations support this idea, as all of them possess identical peaks at low energies, regardless of edge structure and nanoribbon width. We shift the energies of nanoribbons to the left or right, until those peaks are aligned with the corresponding peaks from the DOS of the 2D material. This procedure is shown in Fig. \ref{3}. To make the comparison as accurate as possible, we recalculate the DOS of the single layer using an identical unit cell with the same number of atoms and dimensions, as in the case of nanoribbons with width n$_c$=6, and metal-terminated edge, and periodic boundary conditions along all axes. At low energies, sulfides have three characteristic peaks, a doublet and singlet, while selenides have a triplet and a singlet peak. We chose the shift of energies such that the average error from these peaks is minimum.

 Having aligned the DOS, we can now calculate the absolute position of the  Fermi level of the nanoribbon with respect to the 2D material. In all cases, we find that all nanoribbons have Fermi levels that lie inside the gap of the 2D material. This observation establishes that the metallic character is a universal feature of the zig-zag edges of TMDs nanoribbons. Moreover, these gap states are new states that appeared due to the formation of the edges and are not Bloch electronic states of the single layer. Fig. \ref{3} provides a direct proof that the metallic character of the nanoribbons is due to the collapse of Bloch states into one-dimensional localized electron wave functions along the edges.

 Moreover, the pinning of the Fermi level of the nanoribbons with respect to the single layer is always negative. The presence of metallic phase is accompanied by a decrease of the Fermi level in all cases. For the most stable structures with $N_X=2$ chalcogen adatoms and for a width of $n_c=6$, we observe the same pinning of the Fermi level of about $-0.3$ eV to $-0.5$ eV for all materials.

\subsubsection{Band structure of MoS$_2$ nanoribbons}
\paragraph{} 
Fig. \ref{5} shows the band structures for 2D MoS$_2$ and for nanoribbons of a typical width $n_c=6$. The energies are given relative to the Fermi level  of the 2D material. In all plots, the underlying band structure of the 2D material can be seen. By comparing Figs. \ref{3} and \ref{5}, we see that the valence band maximum of the 2D system is at about -0.75 eV while the conduction band mimimum is at about 0.85 eV, resulting at a band gap of 1.6 eV according to this calculation. The gap is direct, as the two extremes happen at the same $\mathbf{k}$-point which lies at about $\frac{3}{4}$ of the $\Gamma$X line. This point corresponds to the K point of the Brillouin zone for the 2D hexagonal structure, after it has been folded infinite times along the $y$ axis to render it one-dimensional. 

 In all cases, the band gap of the 2D material is crossed by several bands with rich dispersion. Interestingly, the bands are not flat, as one would expect for defect states and localized electrons. These states correspond to truly metallic states of electrons that move in one-dimensional Bloch states along the metallic edge. 

 This universal presence of edge metallic states in a 2D semiconductor, with bands crossing the gap and states that are stable against chemical additions at the edges led to the suggestion  \cite{davelou14} that TMDs are good candidates for topological insulators \cite{kane05}. This observation, if proved correct, would add one more unexpected feature to this  magnificent family of materials.
 
\section{Conclusions}
 We performed a systematic DFT calculation for TMD nanoribbons with different composition (MoS$_2$, MoSe$_2$, WS$_2$ and WSe$_2$), edge structure ($0 \le N_X \le 4$ adatoms at the metal edge) and width ($0 \le n_c \le 7$ or  3.5{\AA} to 42.4{\AA}). We examined the stability of nanoribbons when at equilibrium with chalcogen compounds. We find that nanoribbons with $N_X=2$ adatoms are most stable with respect to solid S or ordinary thiols, while higher number of adatoms is favoured for atomic S or weakly-binded S compounds. Similar results are found for Se compounds. 

 While the 2D materials are semiconductors, TMD nanoribbons with zigzag edges are always metallic regardless of the composition, the width or the edge structures. 
The Fermi level of the metallic phase is always lower in energy than the Fermi level of the 2D material, thus making it favourable for electrons to occupy these metallic states. The bands of the edge states are one-dimensional Bloch states with rich dispersion that cross the gap of the 2D system. This implies that TMD nanoribbons could be prototype systems of 2D topological insulators.

\begin{figure*}
	\begin{tabular}{ccccc}
		\includegraphics[height=3cm]{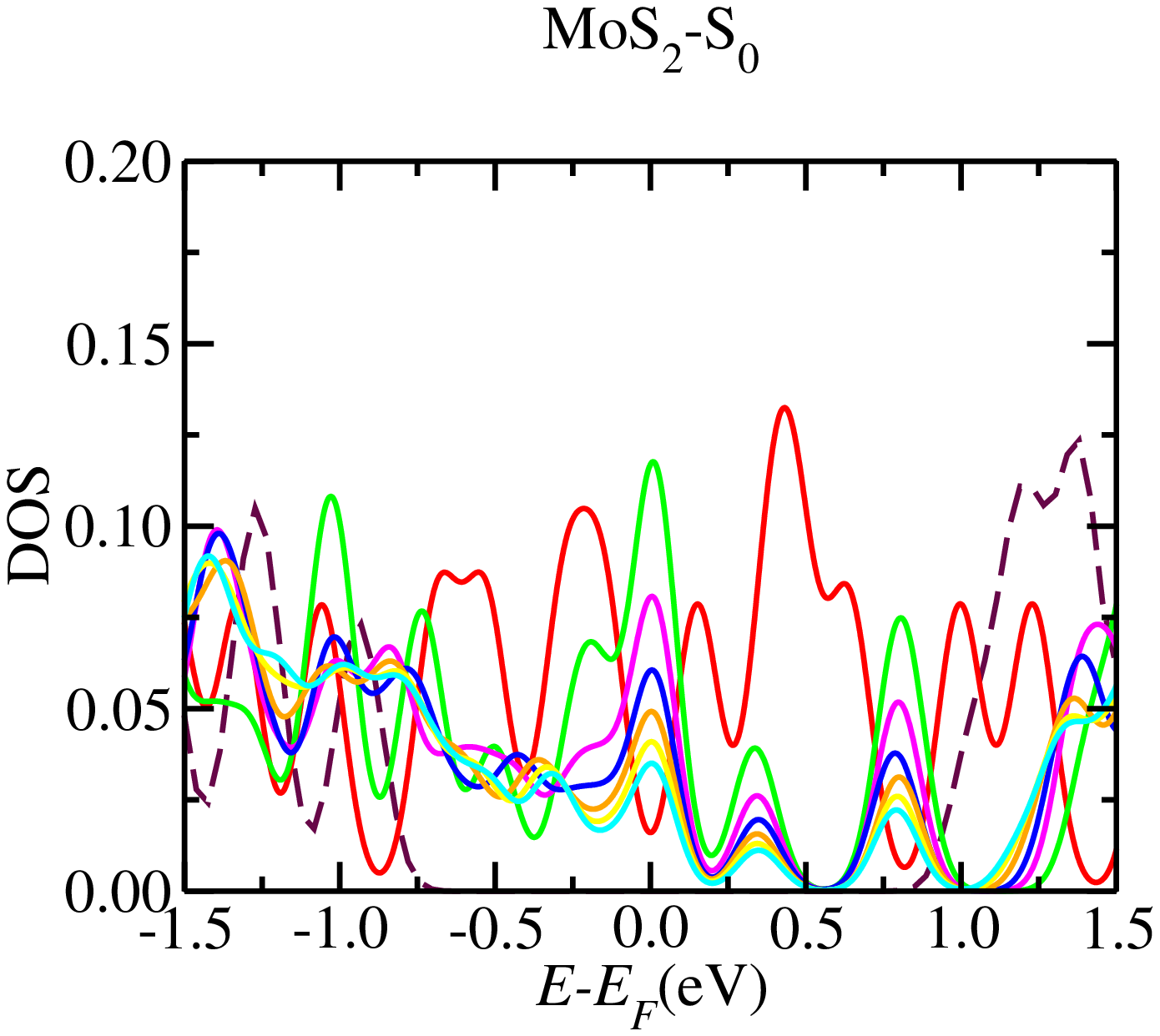} &
		\includegraphics[height=3cm]{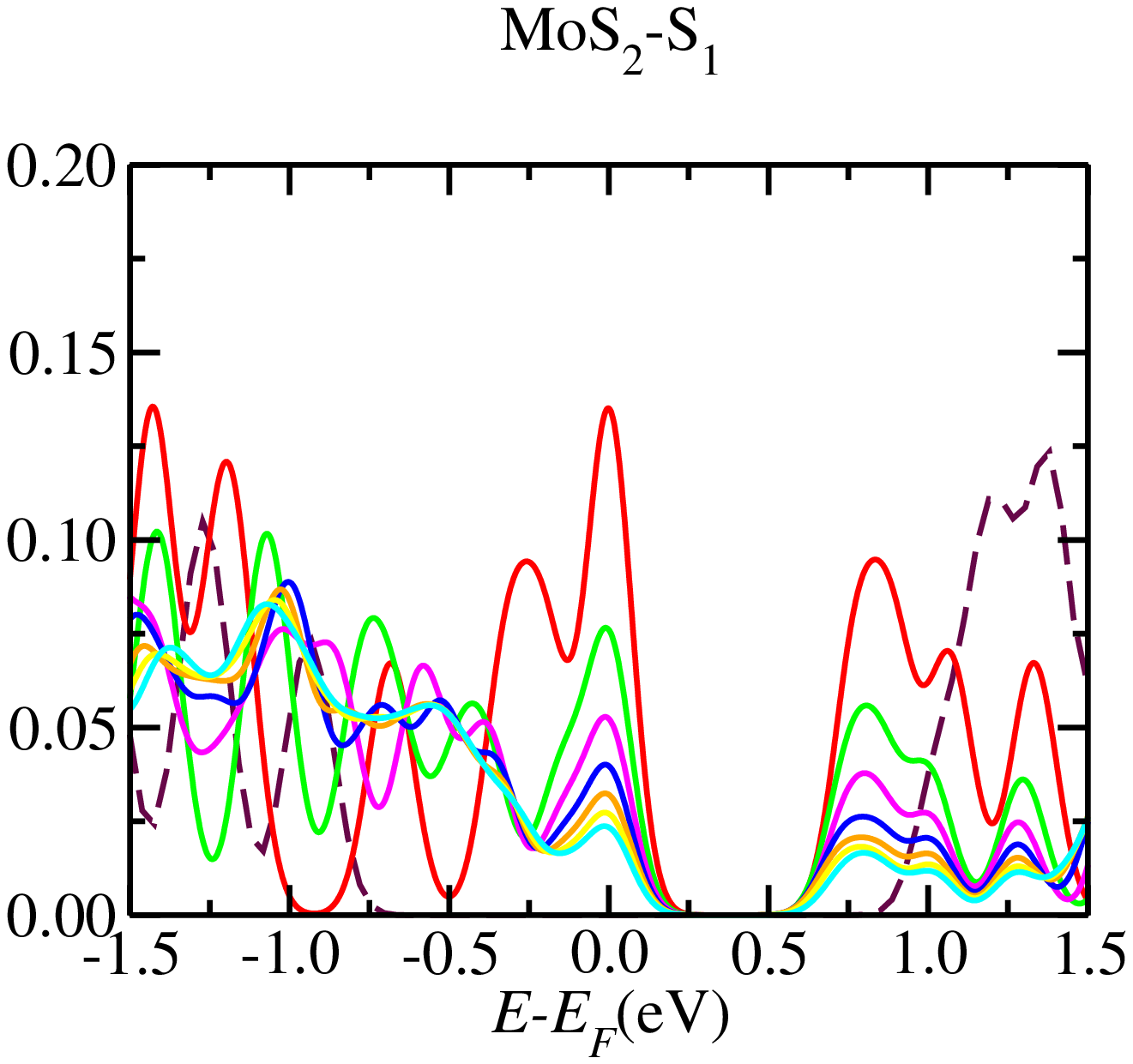} &
		\includegraphics[height=3cm]{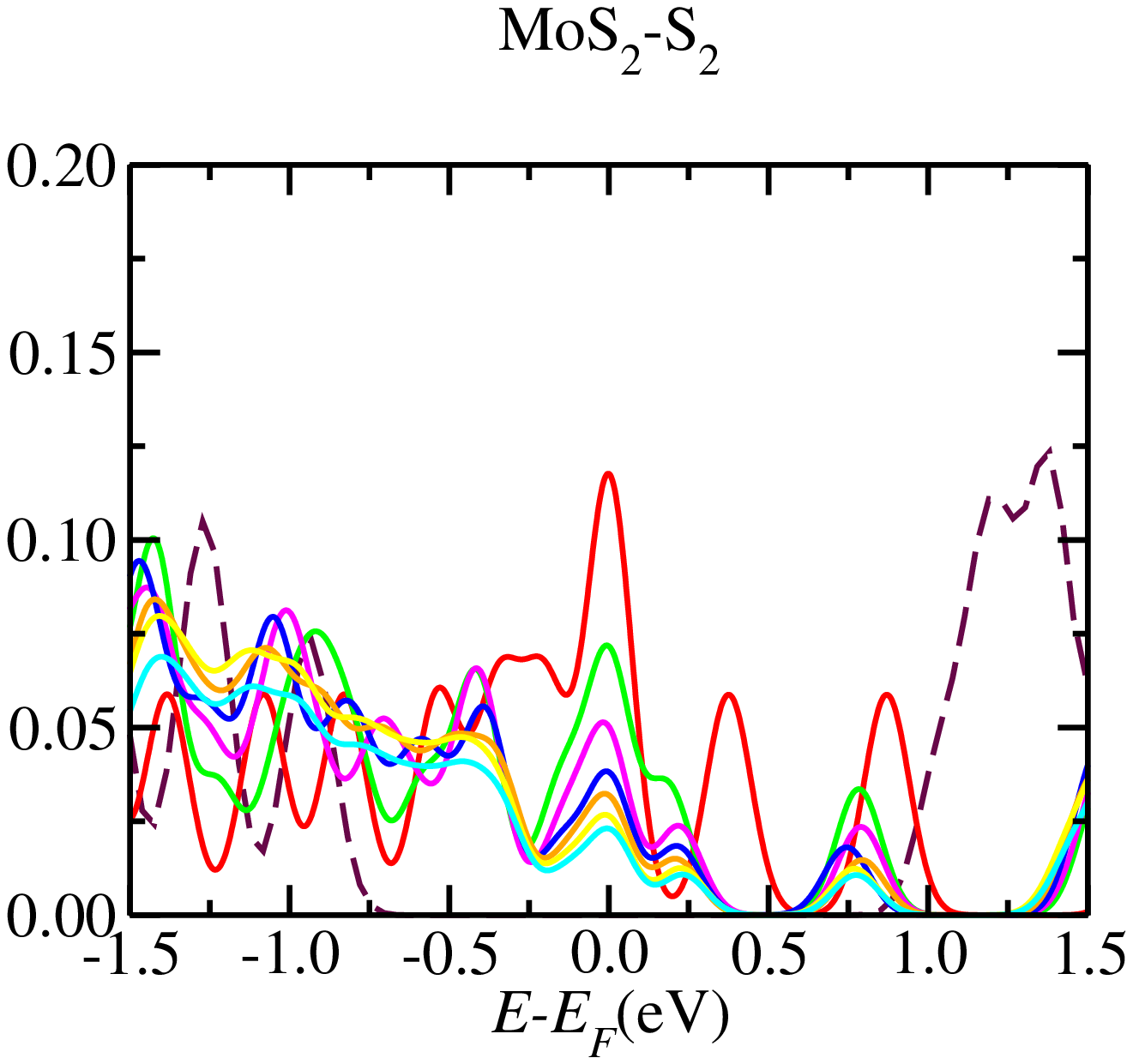} &
		\includegraphics[height=3cm]{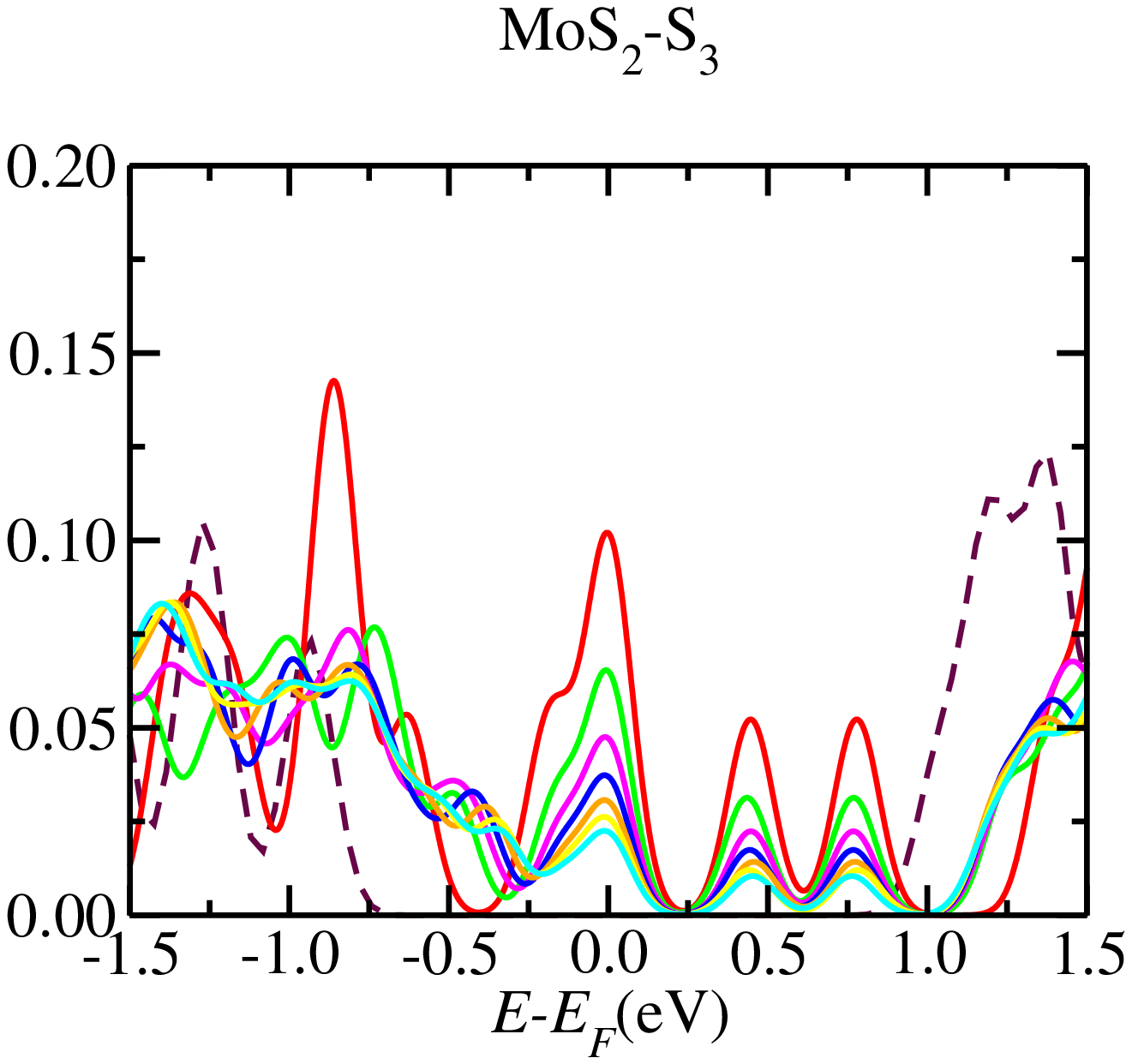} &
		\includegraphics[height=3cm]{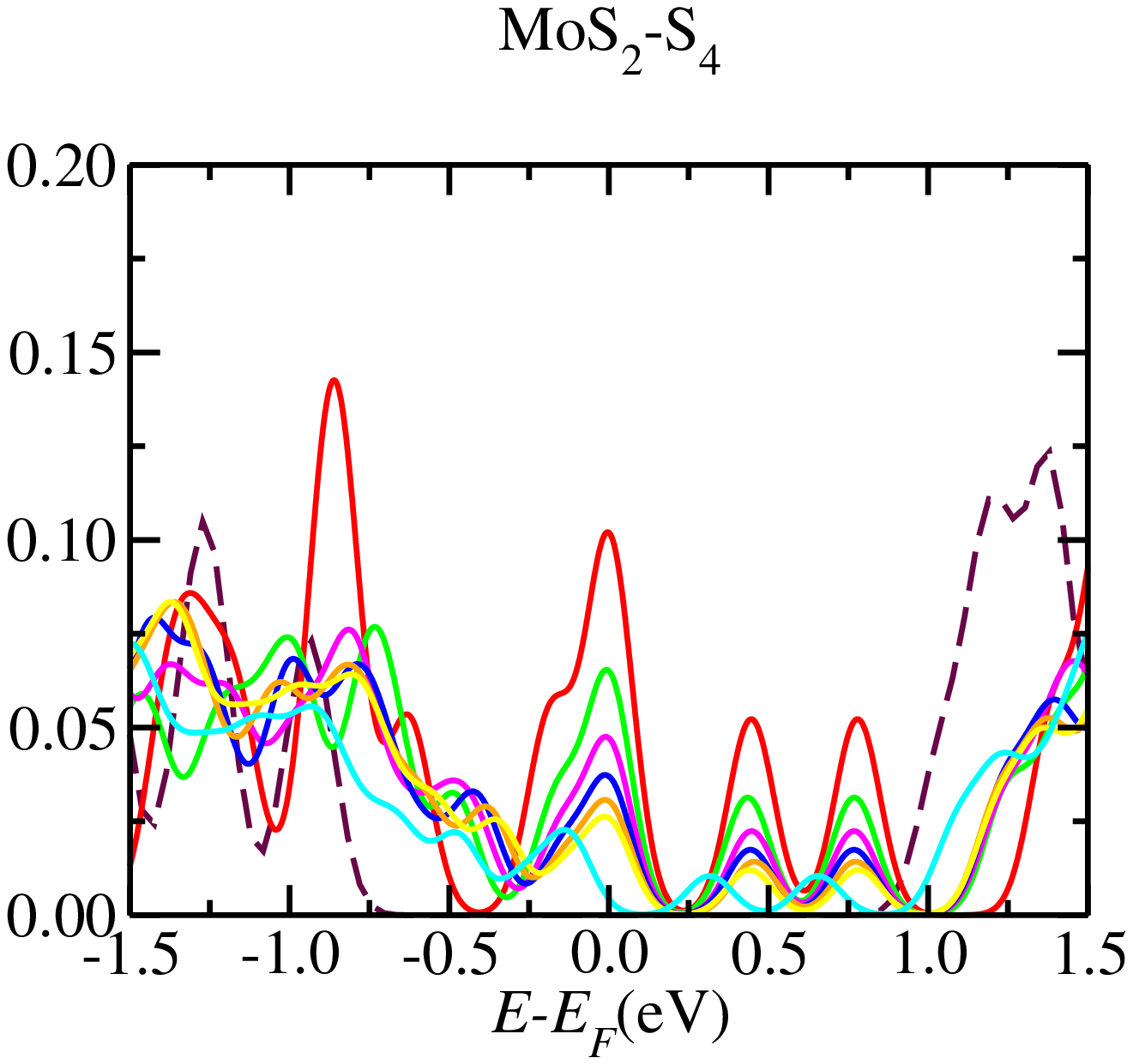} \\
		\includegraphics[height=3cm]{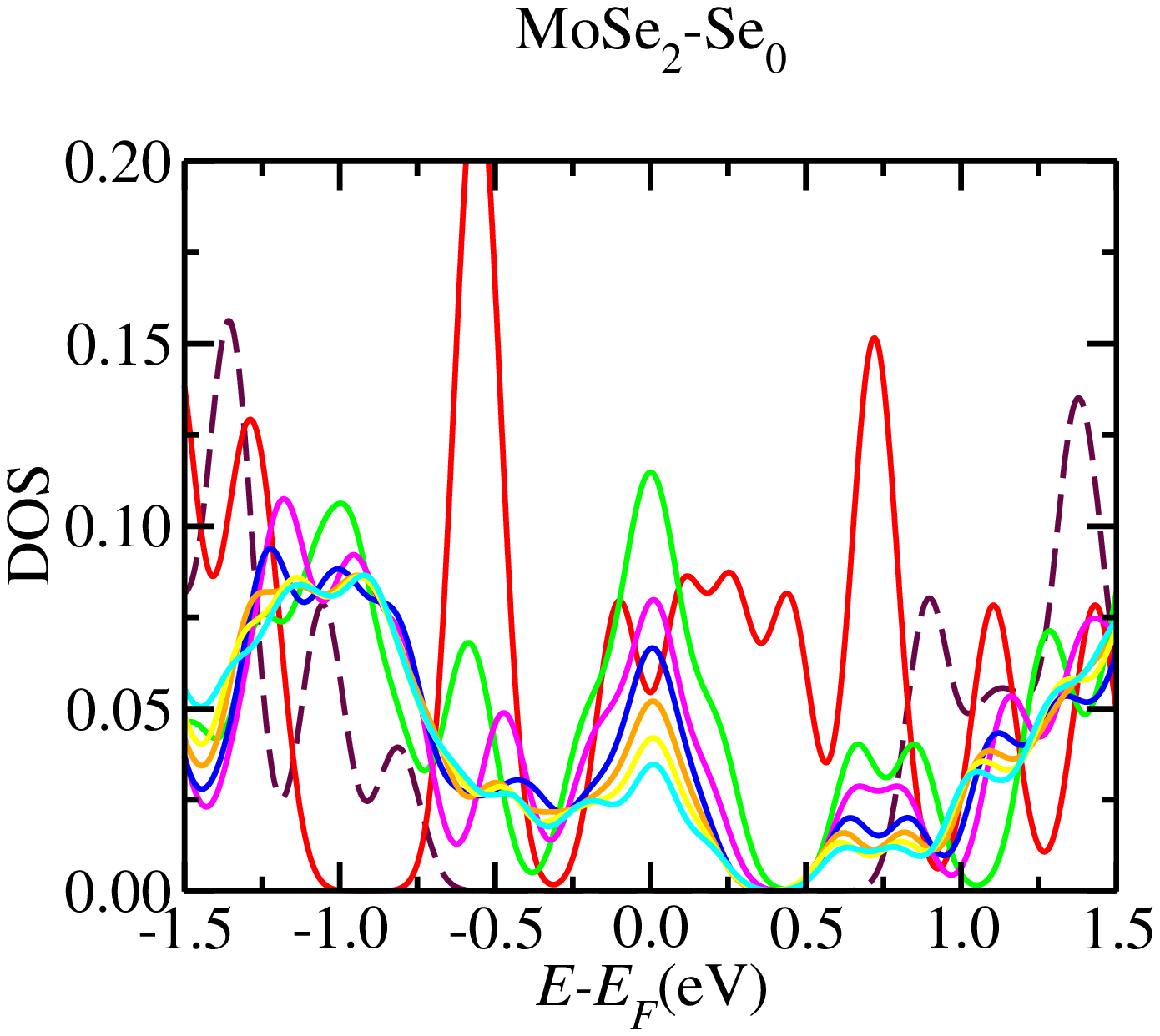} &
		\includegraphics[height=3cm]{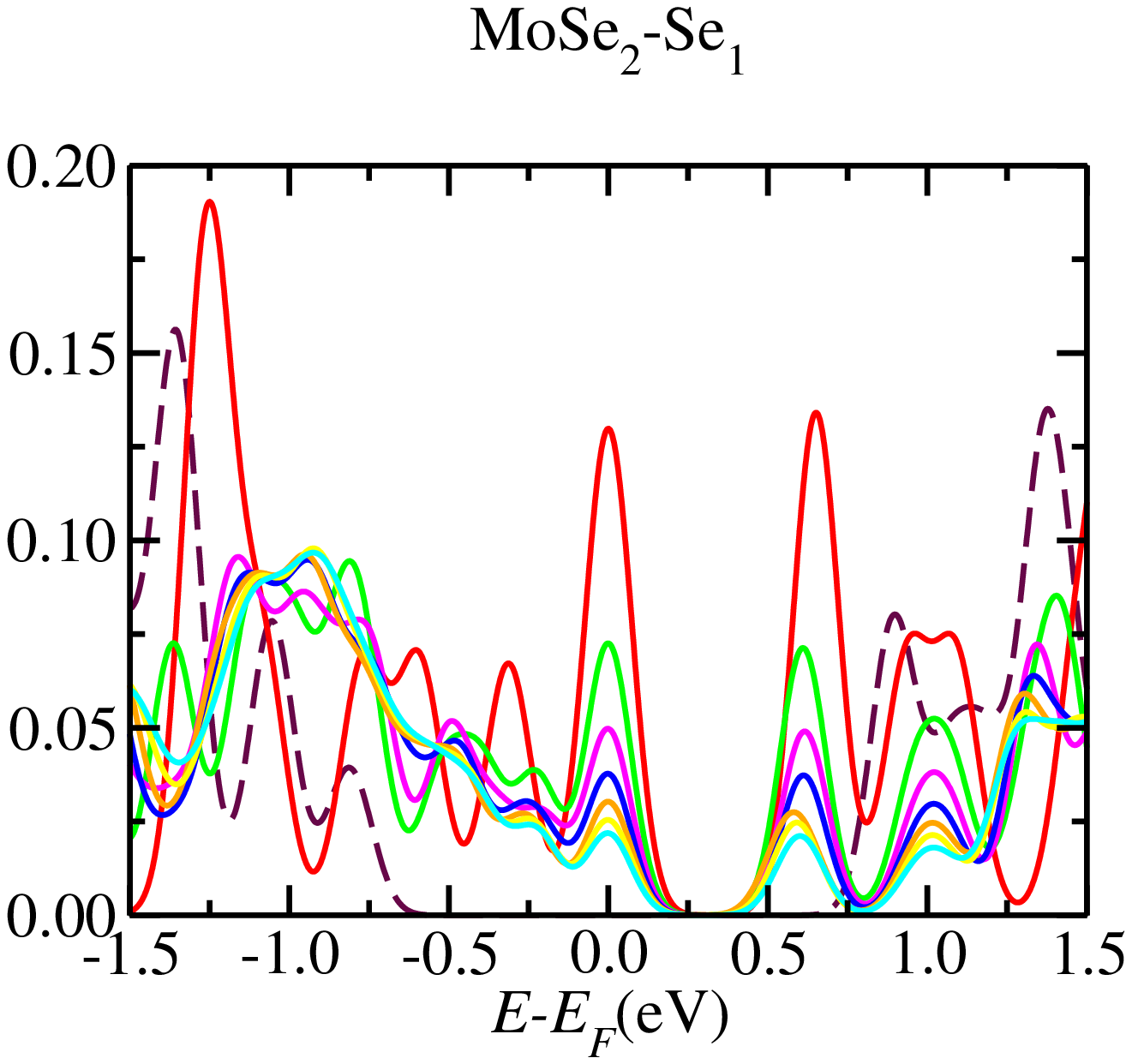} &
		\includegraphics[height=3cm]{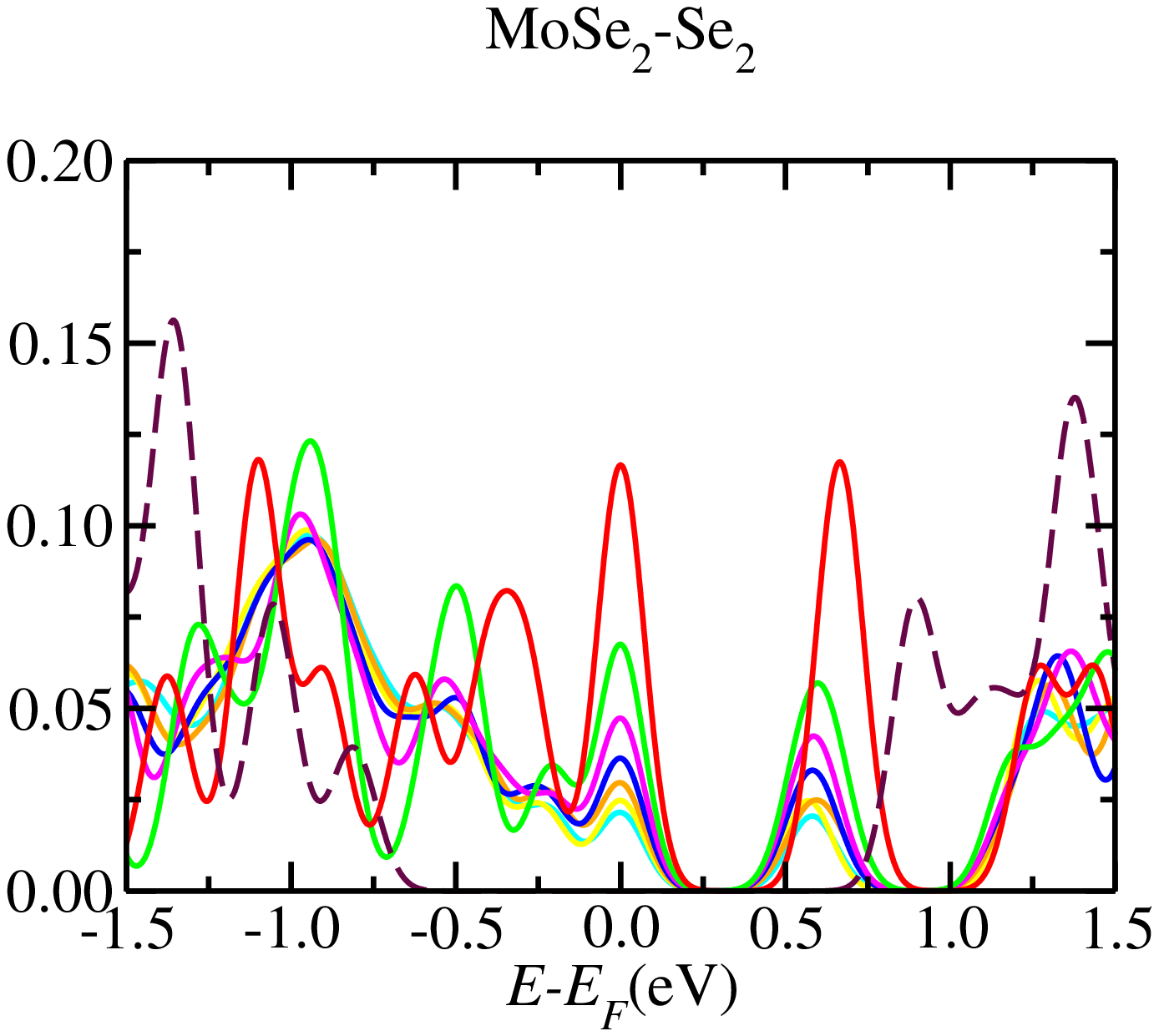} &
		\includegraphics[height=3cm]{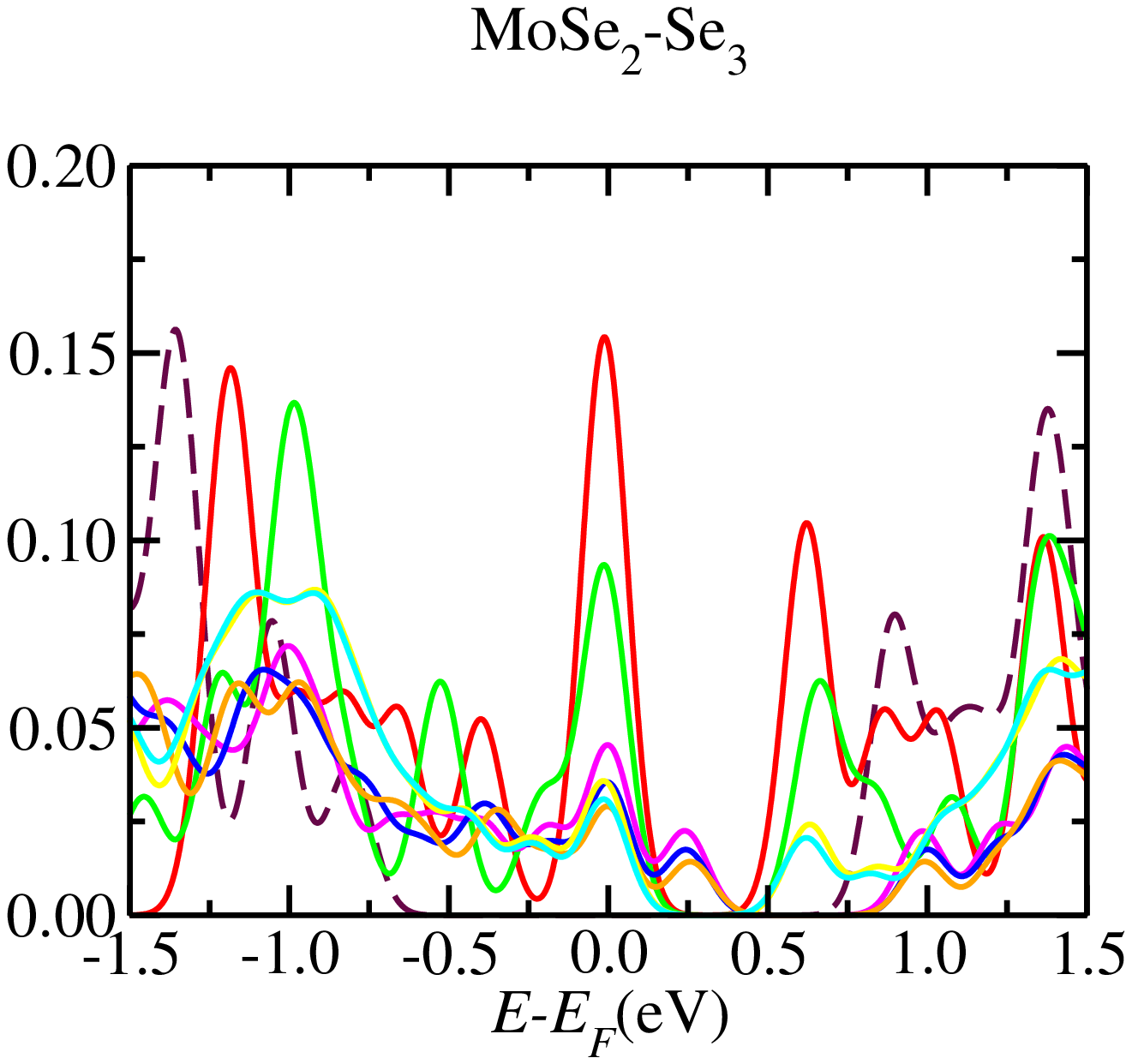} &
		\includegraphics[height=3cm]{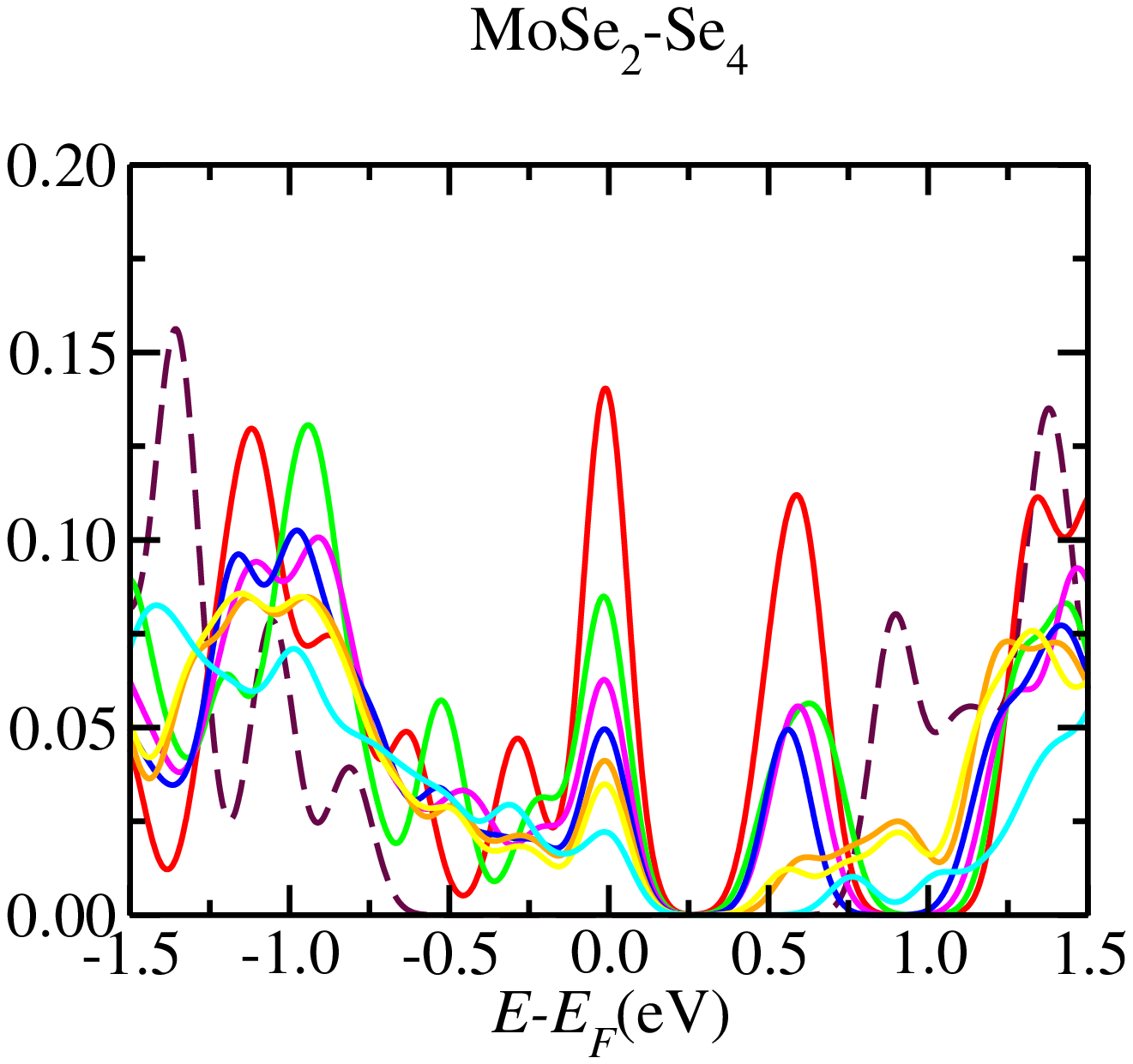} \\
		\includegraphics[height=3cm]{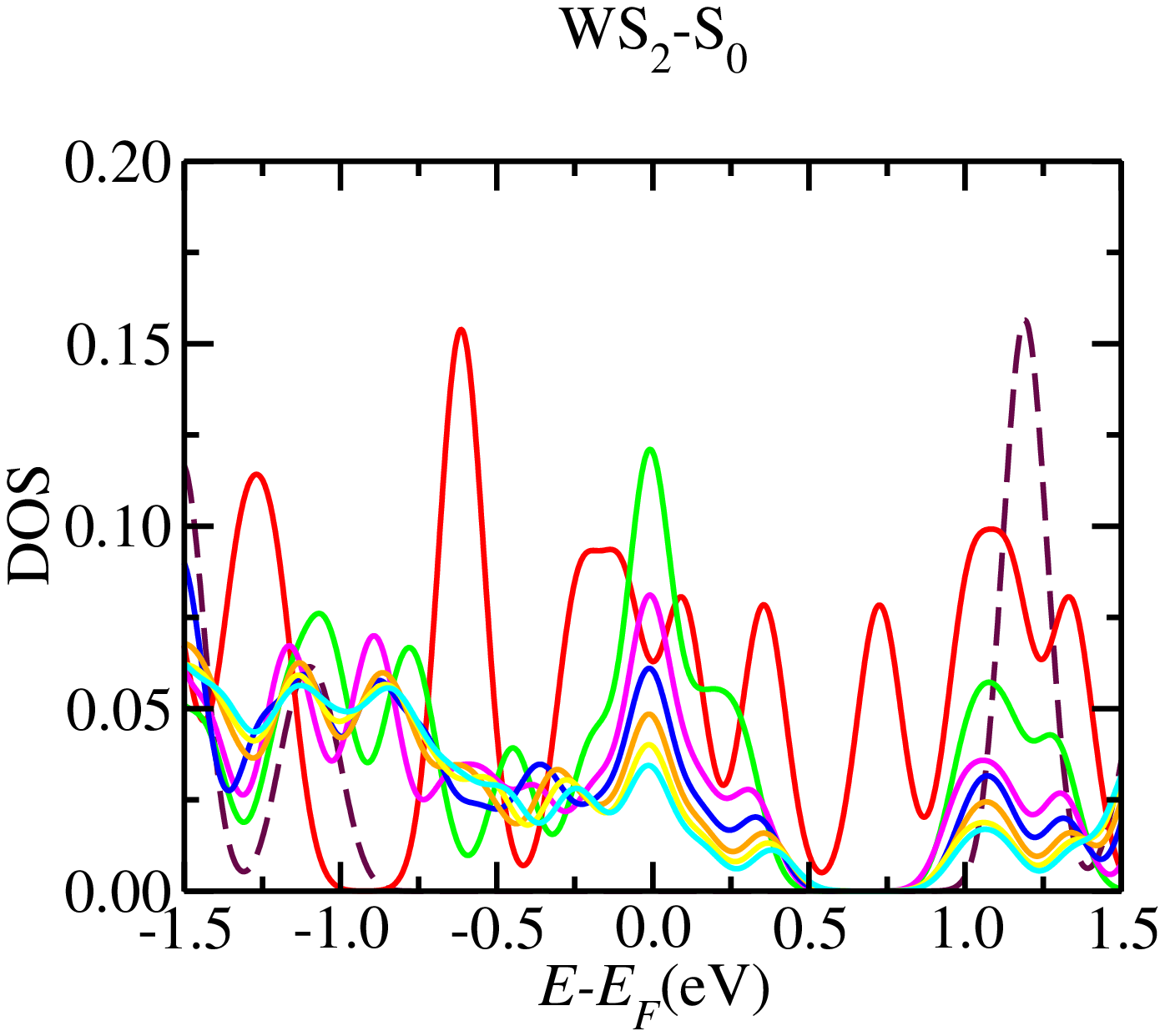} &
		\includegraphics[height=3cm]{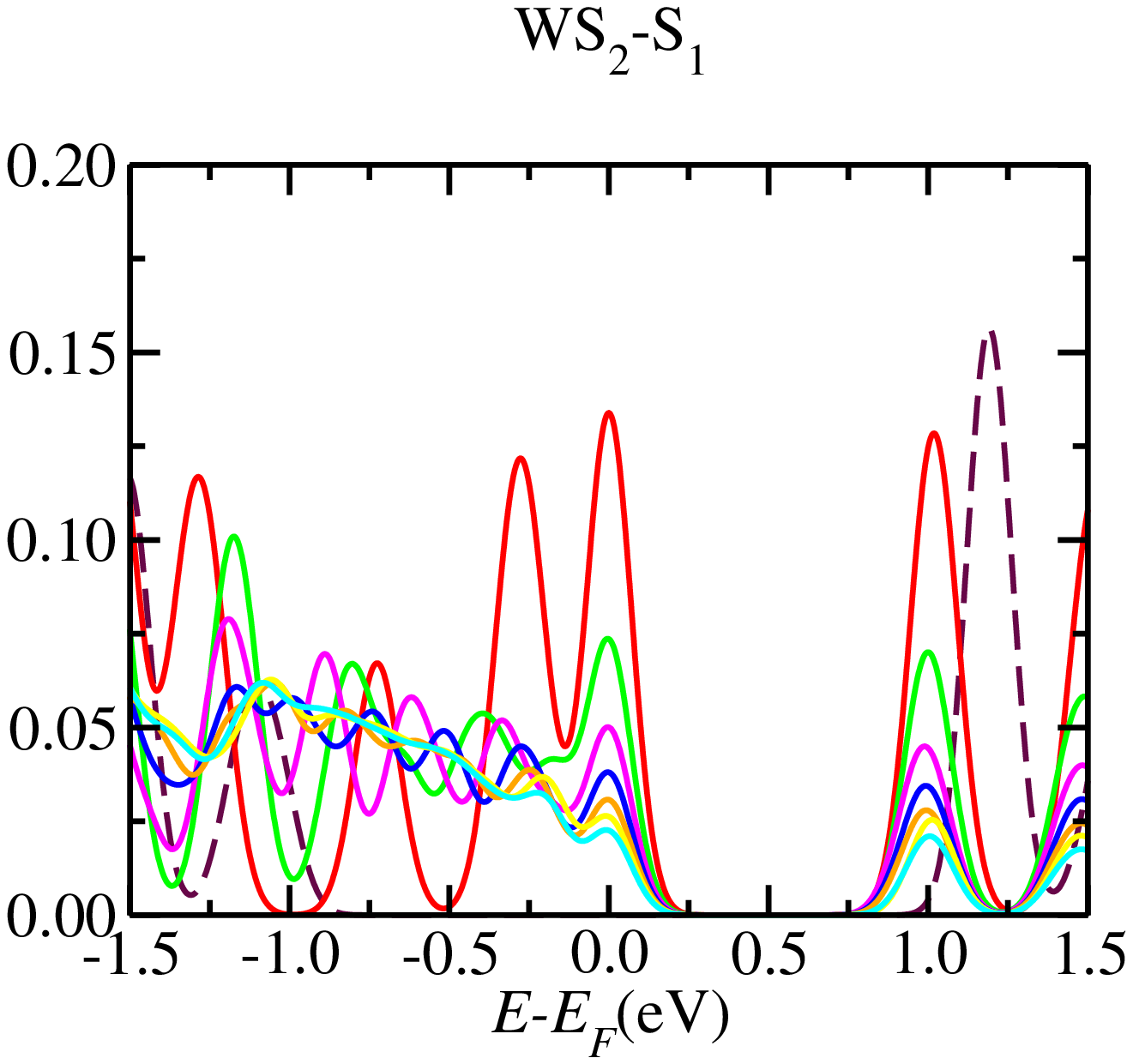} &
		\includegraphics[height=3cm]{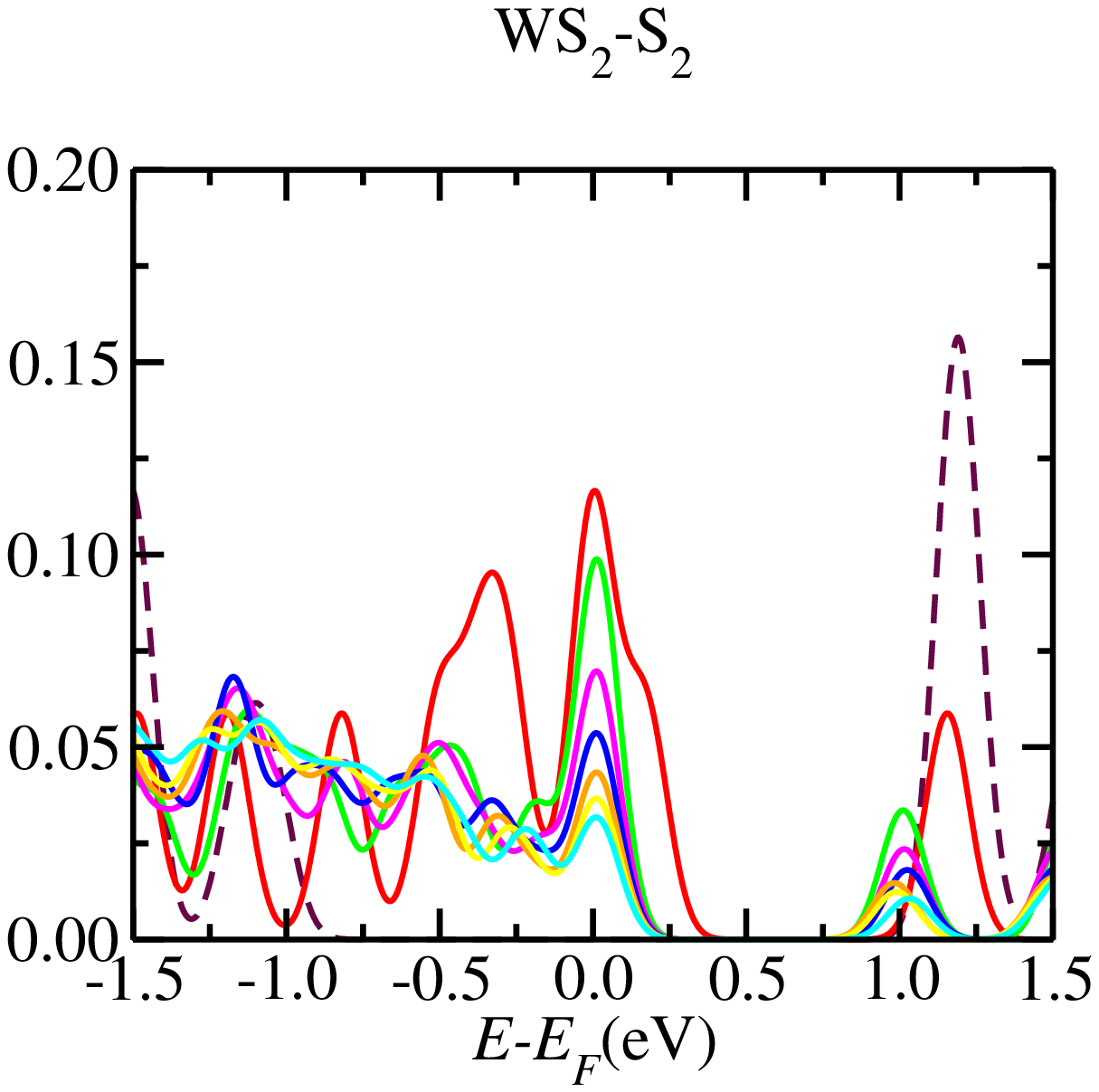} &
		\includegraphics[height=3cm]{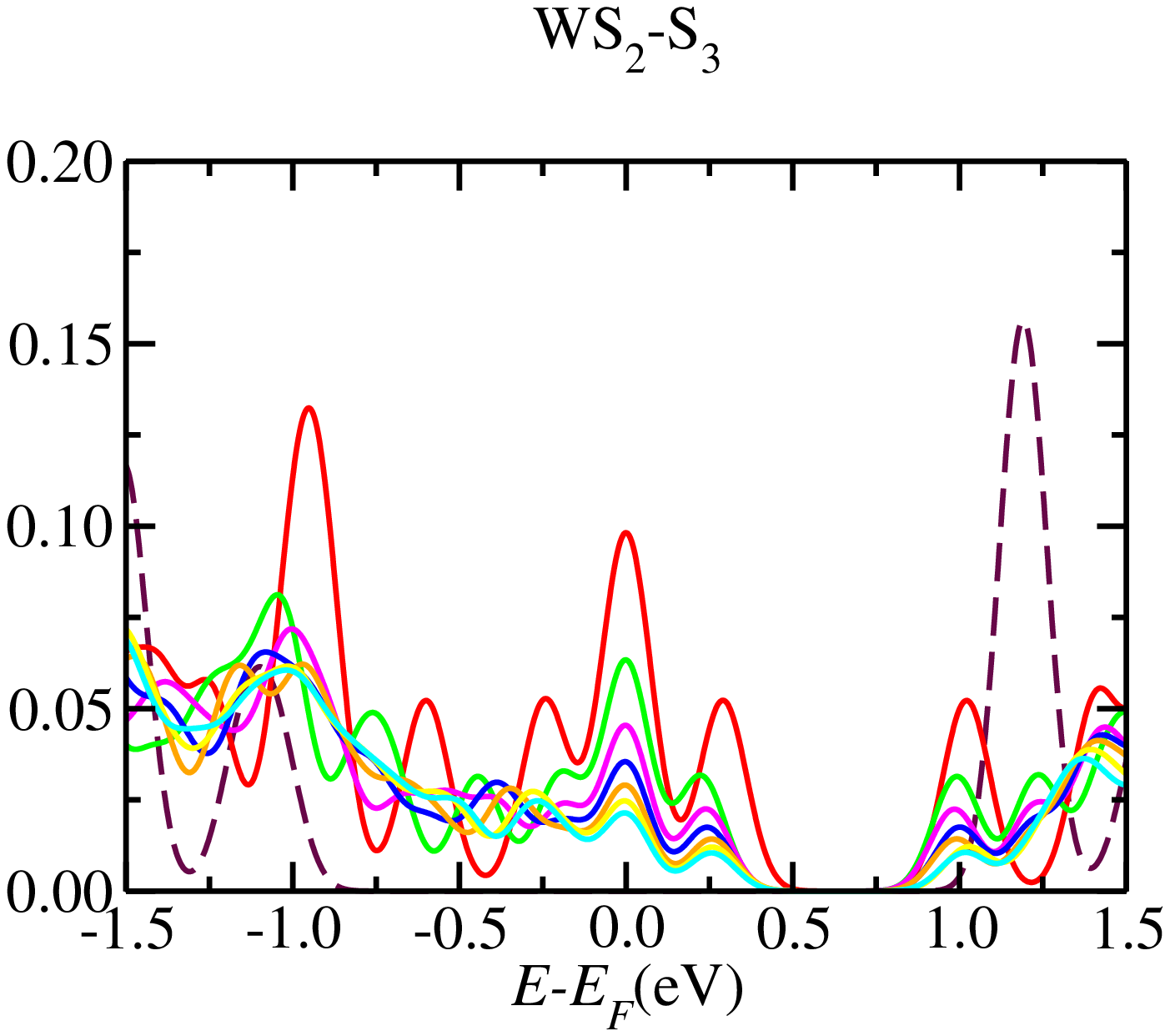} &
		\includegraphics[height=3cm]{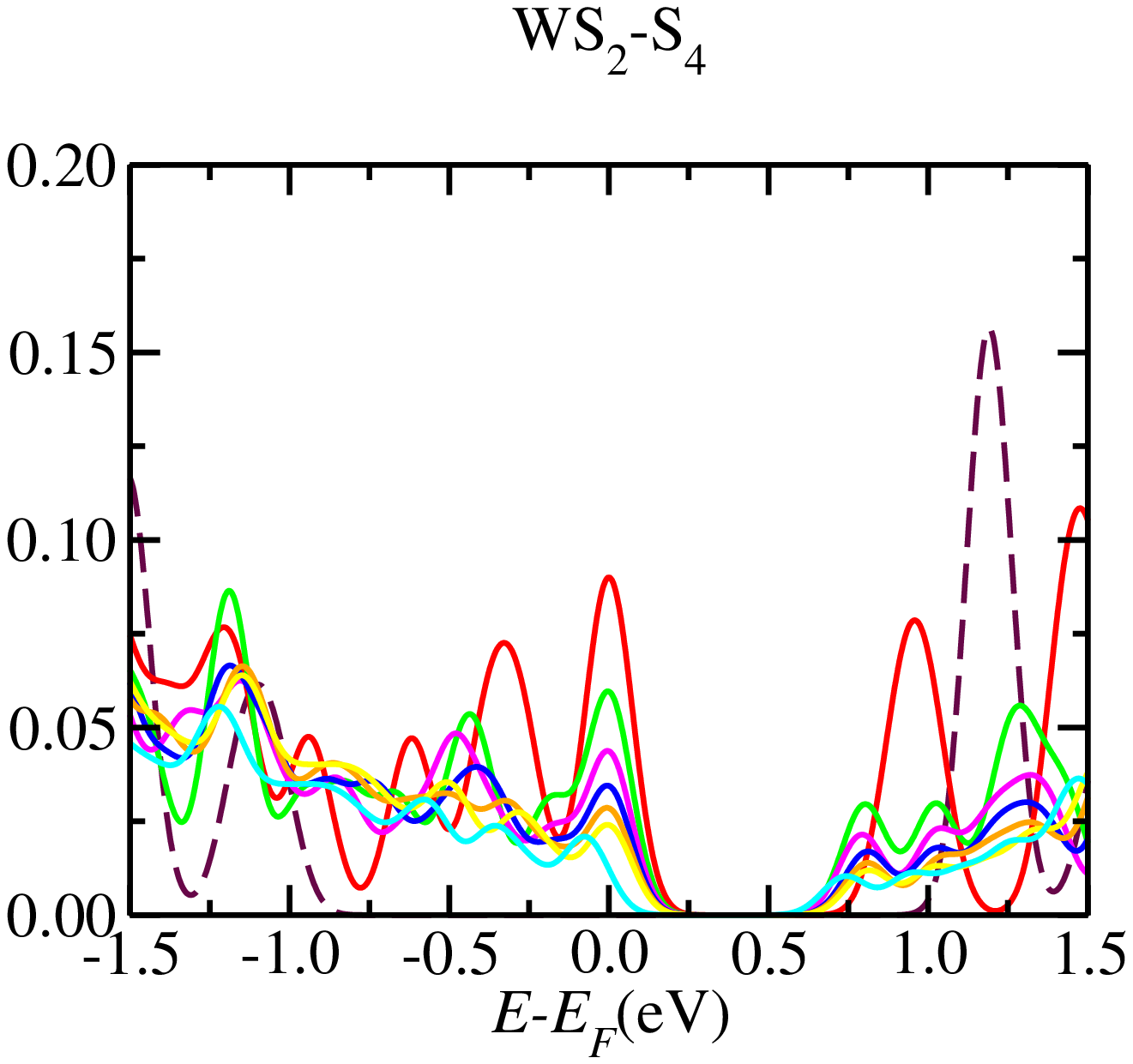} \\
		\includegraphics[height=3cm]{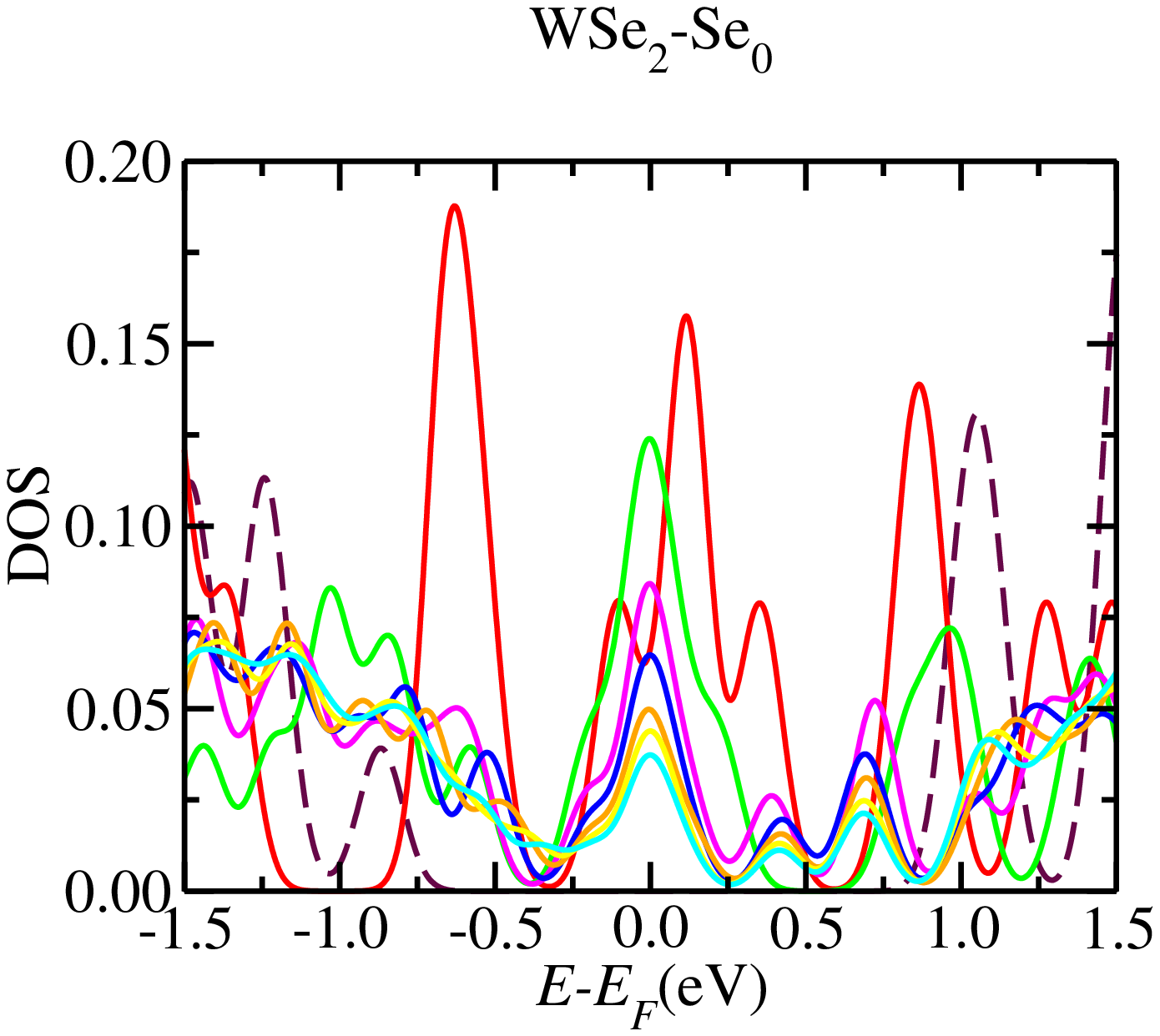} & 
		\includegraphics[height=3cm]{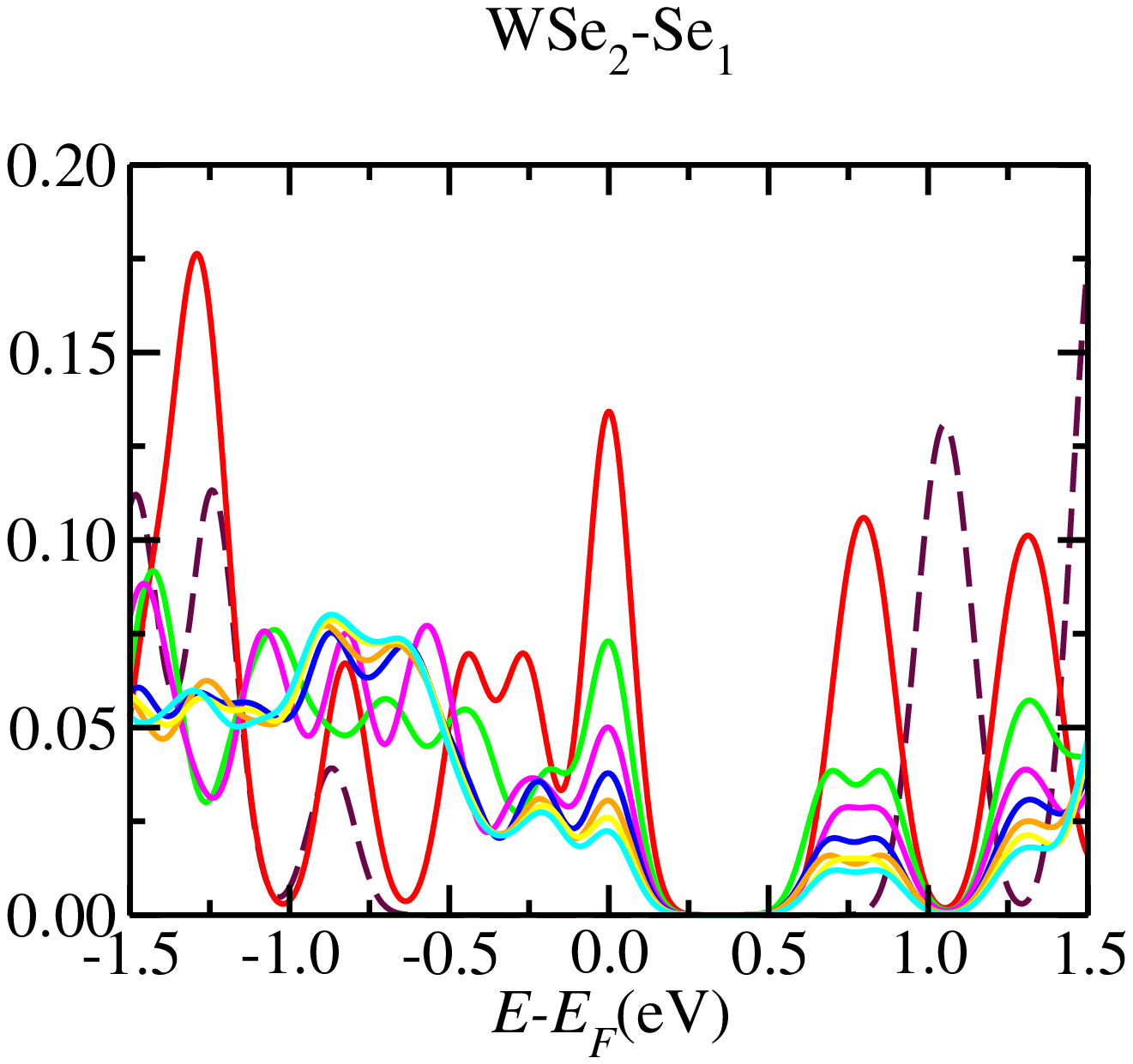} &
		\includegraphics[height=3cm]{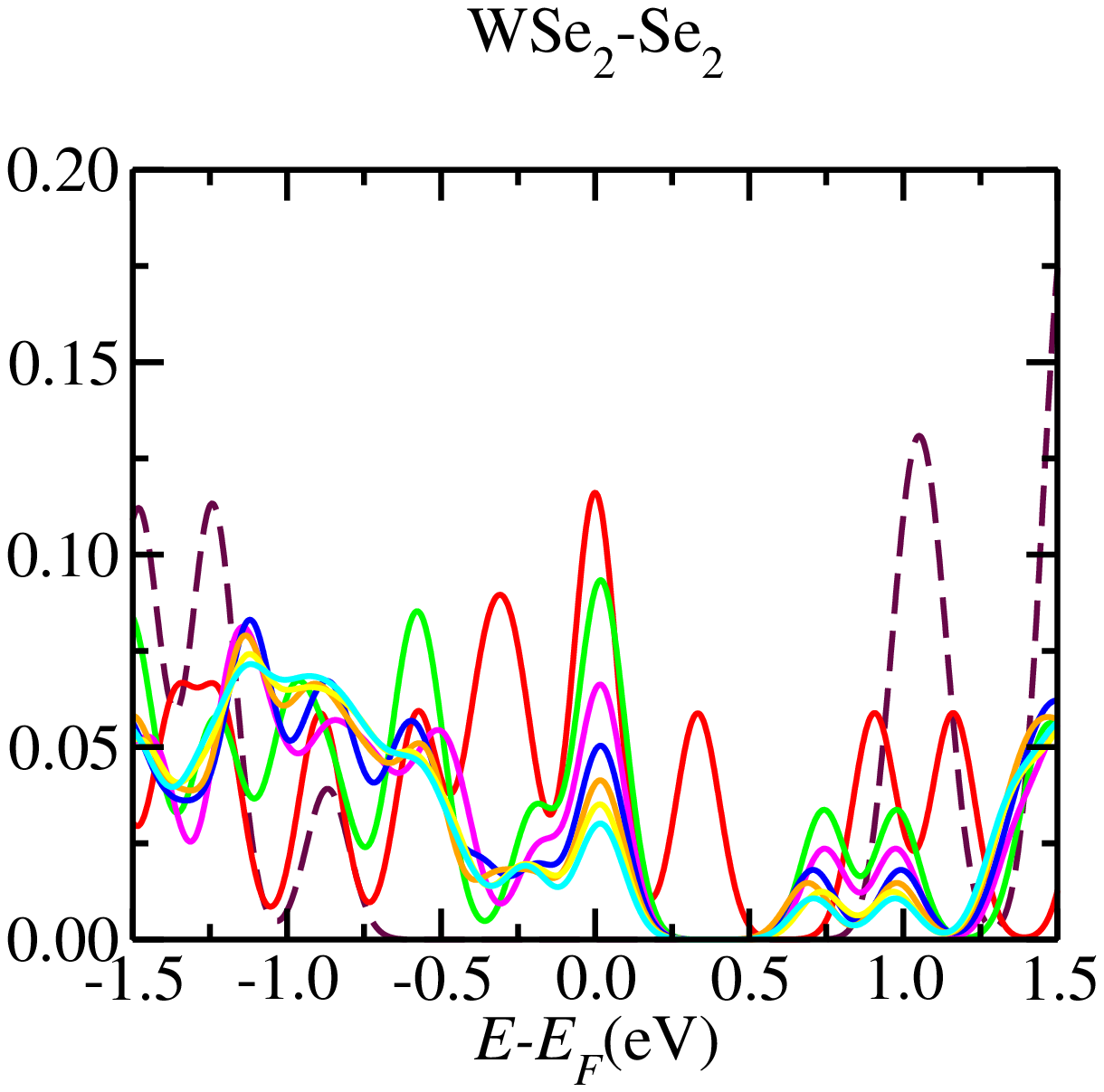} &
		\includegraphics[height=3cm]{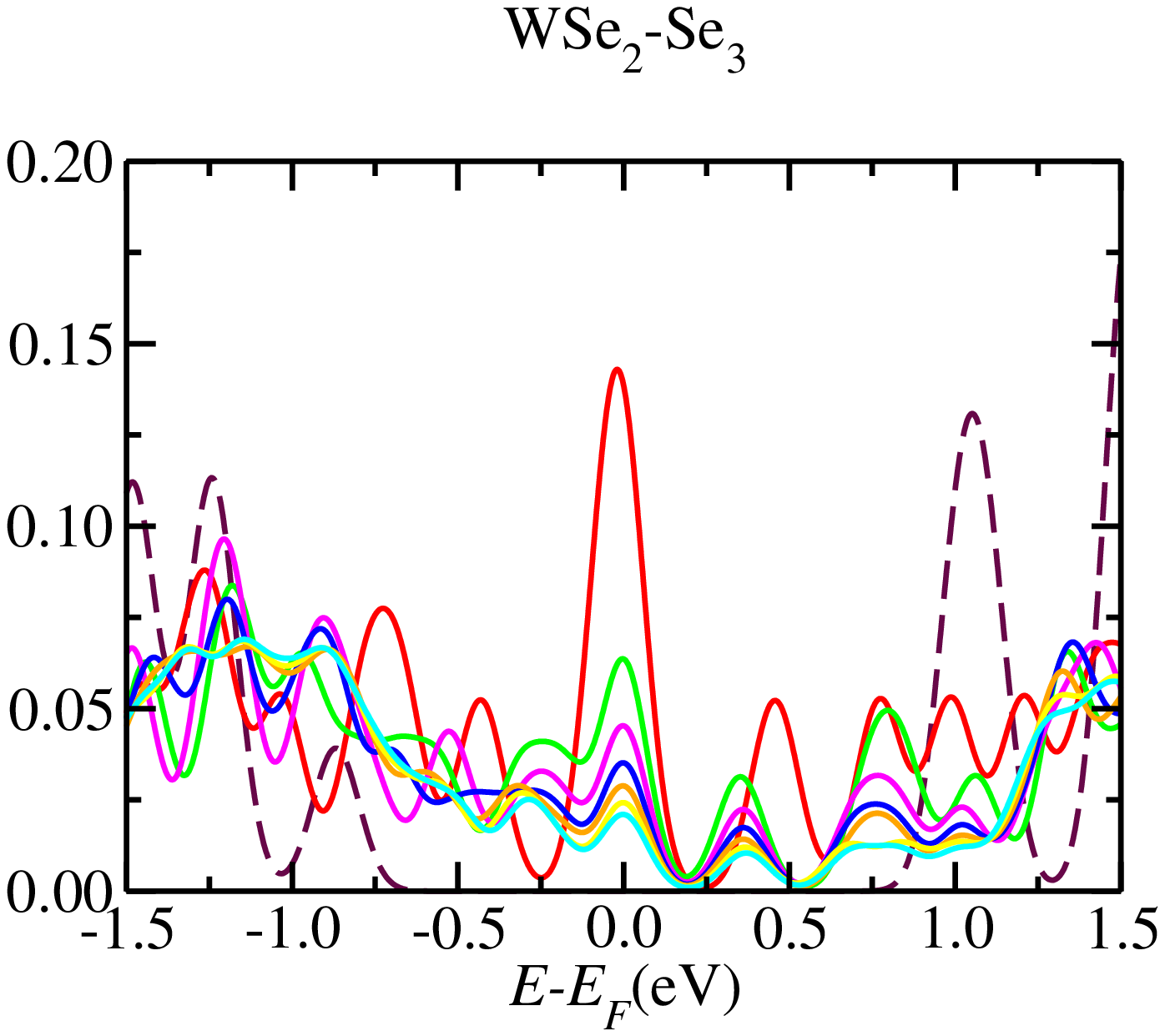} &
		\includegraphics[height=3cm]{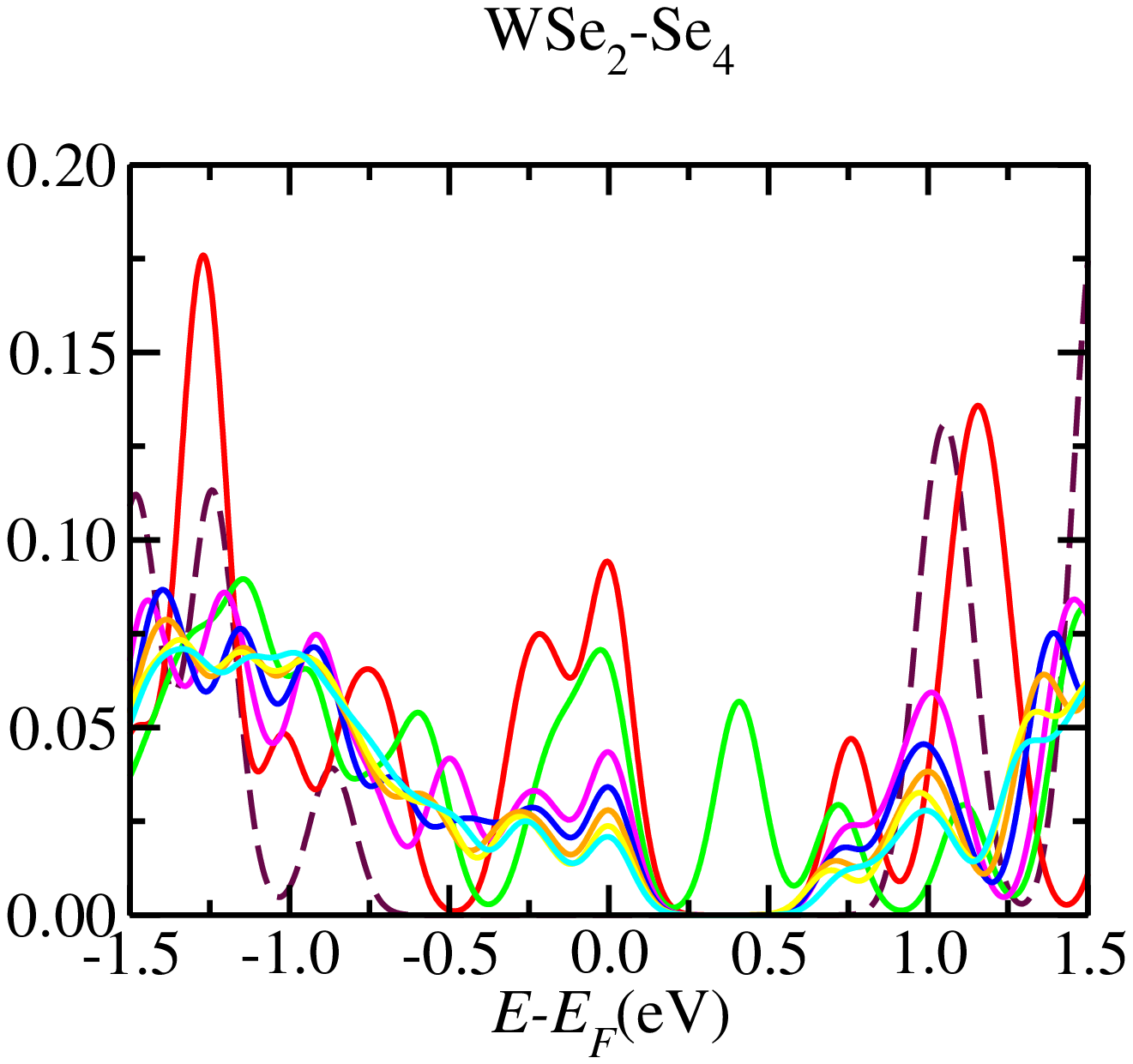} 
	\end{tabular}
	\caption{\label{6}Density of states of TMD nanoribbons with different widths and edge termination.The various colours correspond to the widths of the nanoribbons.The red curve shows the highest intensity at the Fermi level and it represents the width $n_c$=1, while the lowest intensity curve (cyan), represents the width $n_c$=7.From the rest curves we see that the width is reversely proportional to the intensity of the DOS at the Fermi level.}
\end{figure*}

\section{Appendix}
In Fig. \ref{6}, we present a comparison of the density of states of the MX$_2$ nanoribbons with the respective 2D materials. We observe similar features in all calculations. We confirm the presence of edge states at the Fermi level; the qualitative features of these states are not affected by the composition of the material, the different edge termination or the width of the nanoribbon.

\section{Acknowledgements}
\begin{acknowledgements}
This work was supported by the Research Committee of the University of Crete (K.A. 4046 and 4180), by the by Greek GSRT project ERC02-EXEL Grant No. 6260 and by 
EC FET Graphene Flagship.
\end{acknowledgements}

\newpage

\end{document}